\documentclass[%
 twocolumn,aps,pra,amsmath,amssymb,superscriptaddress,nofootinbib,reprint,%
]{revtex4-1}
\pdfoutput=1
\usepackage{stmaryrd}
\usepackage{color}
\usepackage{array}
\usepackage{enumitem}
\usepackage{mathpazo}
\usepackage{setspace}
\usepackage{bm}
\usepackage{amssymb}
\usepackage{amsthm}
\usepackage{mathtools}
\usepackage{multirow}
\usepackage{cases}
\usepackage[colorlinks=true,linkcolor=blue,citecolor=blue,pdfauthor={ },pdftitle={ },pdfsubject={ },pdfkeywords={ }]{hyperref}
\usepackage{pgfplots}
\usepackage{lipsum}
\usepackage{youngtab}
\usepackage{subfigure}

\usepackage[capitalise]{cleveref}

\newtheorem{definition}{Definition}
\newtheorem{proposition}[definition]{Proposition}
\newtheorem{lemma}[definition]{Lemma}

\newtheorem{theorem}[definition]{Theorem}

\newtheorem{corollary}[definition]{Corollary}
\newtheorem{conjecture}[definition]{Conjecture}

\newtheorem{remark}[definition]{Remark}
\newtheorem{example}[definition]{Example}
\newtheorem{question}[definition]{Question}

\def\Dbar{\leavevmode\lower.6ex\hbox to 0pt
{\hskip-.23ex\accent"16\hss}D}
\makeatletter
\def\url@leostyle{%
  \@ifundefined{selectfont}{\def\UrlFont{\sf}}{\def\UrlFont{\small\ttfamily}}}
\makeatother

\DeclareMathOperator{\tr}{tr} %

\def\bcj{\begin{conjecture}}
\def\ecj{\end{conjecture}}
\def\bcr{\begin{corollary}}
\def\ecr{\end{corollary}}
\def\bd{\begin{definition}}
\def\ed{\end{definition}}
\def\bea{\begin{eqnarray}}
\def\eea{\end{eqnarray}}
\def\bem{\begin{enumerate}}
\def\eem{\end{enumerate}}
\def\bex{\begin{example}}
\def\eex{\end{example}}
\def\bim{\begin{itemize}}
\def\eim{\end{itemize}}
\def\bl{\begin{lemma}}
\def\el{\end{lemma}}
\def\bpf{\begin{proof}}
\def\epf{\end{proof}}
\def\bpp{\begin{proposition}}
\def\epp{\end{proposition}}
\def\bqu{\begin{question}}
\def\equ{\end{question}}
\def\br{\begin{remark}}
\def\er{\end{remark}}
\def\bt{\begin{theorem}}
\def\et{\end{theorem}}

\def\btb{\begin{tabular}}
\def\etb{\end{tabular}}

\newcommand{\nc}{\newcommand}

 \nc{\bA}{{\bf A}} \nc{\bB}{{\bf B}} \nc{\bC}{{\bf C}}
 \nc{\bD}{{\bf D}} \nc{\bE}{{\bf E}} \nc{\bF}{{\bf F}}
 \nc{\bG}{{\bf G}} \nc{\bH}{{\bf H}} \nc{\bI}{{\bf I}}
 \nc{\bJ}{{\bf J}} \nc{\bK}{{\bf K}} \nc{\bL}{{\bf L}}
 \nc{\bM}{{\bf M}} \nc{\bN}{{\bf N}} \nc{\bO}{{\bf O}}
 \nc{\bP}{{\bf P}} \nc{\bQ}{{\bf Q}} \nc{\bR}{{\bf R}}
 \nc{\bS}{{\bf S}} \nc{\bT}{{\bf T}} \nc{\bU}{{\bf U}}
 \nc{\bV}{{\bf V}} \nc{\bW}{{\bf W}} \nc{\bX}{{\bf X}}
 \nc{\bZ}{{\bf Z}}

\nc{\cA}{{\cal A}} \nc{\cB}{{\cal B}} \nc{\cC}{{\cal C}}
\nc{\cD}{{\cal D}} \nc{\cE}{{\cal E}} \nc{\cF}{{\cal F}}
\nc{\cG}{{\cal G}} \nc{\cH}{{\cal H}} \nc{\cI}{{\cal I}}
\nc{\cJ}{{\cal J}} \nc{\cK}{{\cal K}} \nc{\cL}{{\cal L}}
\nc{\cM}{{\cal M}} \nc{\cN}{{\cal N}} \nc{\cO}{{\cal O}}
\nc{\cP}{{\cal P}} \nc{\cQ}{{\cal Q}} \nc{\cR}{{\cal R}}
\nc{\cS}{{\cal S}} \nc{\cT}{{\cal T}} \nc{\cU}{{\cal U}}
\nc{\cV}{{\cal V}} \nc{\cW}{{\cal W}} \nc{\cX}{{\cal X}}
\nc{\cZ}{{\cal Z}}

\nc{\hA}{{\hat{A}}} \nc{\hB}{{\hat{B}}} \nc{\hC}{{\hat{C}}}
\nc{\hD}{{\hat{D}}} \nc{\hE}{{\hat{E}}} \nc{\hF}{{\hat{F}}}
\nc{\hG}{{\hat{G}}} \nc{\hH}{{\hat{H}}} \nc{\hI}{{\hat{I}}}
\nc{\hJ}{{\hat{J}}} \nc{\hK}{{\hat{K}}} \nc{\hL}{{\hat{L}}}
\nc{\hM}{{\hat{M}}} \nc{\hN}{{\hat{N}}} \nc{\hO}{{\hat{O}}}
\nc{\hP}{{\hat{P}}} \nc{\hR}{{\hat{R}}} \nc{\hS}{{\hat{S}}}
\nc{\hT}{{\hat{T}}} \nc{\hU}{{\hat{U}}} \nc{\hV}{{\hat{V}}}
\nc{\hW}{{\hat{W}}} \nc{\hX}{{\hat{X}}} \nc{\hZ}{{\hat{Z}}}

\newcommand{\bra}[1]{\langle#1|}
\newcommand{\ket}[1]{|#1\rangle}

\def\Dbar{\leavevmode\lower.6ex\hbox to 0pt
{\hskip-.23ex\accent"16\hss}D}

\def\be{\begin{eqnarray}}
\def\ee{\end{eqnarray}}

\begin{document}
\title{Supervised learning in Hamiltonian reconstruction from local measurements on eigenstates}

\author{Chenfeng Cao}
\email[]{chenfeng.cao@connect.ust.hk}
\affiliation{Department of Physics, The Hong Kong University of Science and Technology, Clear Water Bay, Kowloon, Hong Kong, China}

\author{Shi-Yao Hou}
\affiliation{College of Physics and Electronic Engineering, Center for Computational Sciences,  Sichuan Normal University, Chengdu 610068, China}
\affiliation{Department of Physics, The Hong Kong University of Science and Technology, Clear Water Bay, Kowloon, Hong Kong, China}

\author{Ningping Cao}
\affiliation{Department of Mathematics \& Statistics, University of
	Guelph, Guelph, Ontario, Canada}%
\affiliation{Institute for Quantum Computing, University of Waterloo, Waterloo, Ontario, Canada}

\author{Bei Zeng}
\email[]{zengb@ust.hk}
\affiliation{Department of Physics, The Hong Kong University of Science and Technology, Clear Water Bay, Kowloon, Hong Kong, China}


\date{\today}

\begin{abstract}
Reconstructing a system Hamiltonian through measurements on its eigenstates is an important inverse problem in quantum physics. Recently, it was shown that generic many-body local Hamiltonians can be recovered by local measurements without knowing the values of the correlation functions. In this work, we discuss this problem in more depth for different systems and apply the supervised learning method via neural networks to solve it. For low-lying eigenstates, the inverse problem is well-posed, neural networks turn out to be efficient and scalable even with a shallow network and a small data set. For middle-lying eigenstates, the problem is ill-posed, we present a modified method based on transfer learning accordingly. Neural networks can also efficiently generate appropriate initial points for numerical optimization based on the BFGS method.
\end{abstract}

\maketitle
\renewcommand\theequation{\arabic{section}.\arabic{equation}}
\setcounter{tocdepth}{4}
\makeatletter
\@addtoreset{equation}{section}
\makeatother

\section{Introduction}

Naturally arising physical systems exhibit local interactions. Consequently, their ground and thermal states are uniquely determined by their local marginals~\cite{zeng2015quantum}.  For a many-body quantum system in thermal equilibrium, information from measuring local observables suffices to reconstruct its quantum state~\cite{swingle2014reconstructing,chen2012correlations,bairey2019learning,qi2019determining}. As a comparison, determining a generic pure state will need measurements on subsystems that are half of the system size~\cite{linden2002almost,baldwin2016strictly,huang2018quantum,karuvade2018generic}. Many algorithms are proposed for thermal state reconstruction from local measurements~\cite{zhou2008irreducible, niekamp2013computing}. Experiments have also been performed to demonstrate that the reconstruction is robust against real-world noise~\cite{xin2019local}. 

It is also realized that nondegenerate eigenstates inherit some properties of thermal states. It has been long conjectured that nondegenerate eigenstates of a local Hamiltonian are in fact eigenstates of some other local Hamiltonians~\cite{cioslowski2000many,mazziotti1998contracted}.  This conjecture has been examined from various aspects, including eigenstate correlation~\cite{qi2019determining, chen2012correlations} and the eigenstate thermalization hypothesis~\cite{deutsch1991quantum, srednicki1994chaos, gogolin2016equilibration, garrison2018does}. This conjecture is also known to be closely related to the quantum marginal problem and correlations in many-body systems~\cite{coleman2000reduced}. 

To be more precise, consider a $k$-local Hamiltonian $H = \sum_i c_{i} A_{i}$ with $A_i$s being $k$-local operators acting non-trivially on at most $k$ particles. For any thermal state $\rho$ of the system with temperature known, information of $k$-particle reduced density matrices ($k$-RDMs) suffices to infer $c_{i}$s (hence to infer $H$ and $\rho$). The question now is, for an eigenstate $\ket{\psi}$ of $H$ satisfying $H\ket{\psi}=\lambda\ket{\psi}$ for eigenvalue $\lambda$, whether $k$-RDMs of $\ket{\psi}\bra{\psi}$ would be enough to infer $c_{i}$s. This certainly cannot be true in general as it is easy to construct counterexamples. Surprisingly, as recently shown in~\cite{qi2019determining, hou2020determining}, this is indeed true in generic cases.

Moreover, in Ref.~\cite{hou2020determining}, a method for reconstructing $c_{i}$s is proposed. The method uses only the local measurement information of $\bra{\psi}A_i\ket{\psi}$, which is arguably the minimum possible information to determine $c_{i}$ as one would hope for.  As a comparison, the method discussed in Ref.~\cite{qi2019determining} uses also the correlation information given by $\bra{\psi}A_iA_j\ket{\psi}$.
The method in Ref.~\cite{hou2020determining} transforms the problem into an unconstrained optimization problem.  It is then natural to use the Broyden-Fletcher-Goldfarb-Shanno (BFGS) algorithm~\cite{bonnans2006numerical} to carry out the optimization. Numerical experiments have demonstrated the effectiveness and robustness of the method, which does converge to the desired result. However, this BFGS is mainly based on the Monte Carlo method to search for the initial point, which is easily trapped in a local minimum. The demonstrated performance of the algorithm is very time-consuming, which is at the cost of extensive initial point sampling, and there is no guarantee of convergence. It is hence highly desired to find other methods to approach the problem more efficiently.

In this work, we propose to address the problem from the perspective of an inverse problem, with which lots of machine learning techniques can then be naturally applied.
We notice that the problem of reconstructing $c_{i}$s from $k$-RDM information is a typical inverse problem, which requires calculating the causal factors from observables. In its general form, 
for a deterministic forward model $\mathbf{y}=\mathcal{A}\left(x^{*}\right)+\mathbf{e}$,  where $x^{*}$ is the system parameter, $\mathbf{y}$ is the measured data and $\mathbf{e}$ is the observation noise, given data $\mathbf{y}$, an inverse problem is that we want to recover the model parameter $x^{*}$ from the given data $\mathbf{y}$. In our case, $\mathbf{y}$ is the measurement data given by $\bra{\psi}A_i\ket{\psi}$, and $x^{*}$ are the system parameters given by $c_{i}$s, and $\mathbf{e}$ is the measurement noise. 

Neural network is an efficient approach to approximate the solution for various inverse problems~\cite{adler2017solving}, such as image reconstruction~\cite{mccann2017convolutional, lucas2018using}, signal recovery~\cite{mousavi2017learning}, and learning PDE models from data~\cite{long2017pde}.  We will then address the reconstruction problem with the supervised learning method based on neural network techniques. Notice that 
in general, most inverse problems are ill-posed. From the method based on Monte-Carlo sampling in Ref.~\cite{hou2020determining},  we understand that if $ | \psi_{n} \rangle$ is a low-lying excited state of $H$, there are very few $\{a_{i}\}$s that correspond to multiple $\{c_{i}\}$s, then the Hamiltonian recovery problem is well-posed. When $ | \psi_{n}  \rangle $ is middle-lying ($n \approx 2^{N-1}$ for $N$-qubit system), the solution is highly sensitive to the $k$-RDMs, the problem is therefore ill-posed.

We demonstrate that for low-lying excited states where the inverse problem is well-posed, our new method returns $c_{i}$s with fast speed and high fidelity, compared to the BFGS method. Furthermore,  when the fidelity between the predicted Hamiltonian and the real Hamiltonian is not good enough,  the numerical algorithm in Ref. \cite{hou2020determining} can serve as a supplement step to improve it further. We also show that our method is robust against the noise $\mathbf{e}$. We believe that this method will add new tools to understand the related quantum state inference problem by bringing machine learning techniques from the study of inverse problems, as well as shed new light on the mystery of the eigenstates correlation problem.

The paper is organized as follows: In Sec. II, we discuss the Hamiltonian reconstruction problem for eigenstates and provide some intuition on the uniqueness argument. In Sec. III, we formulate our problem in terms of an inverse problem, present our method based on supervised learning via neural networks, then discuss the results and applications. 

Some discussions on dealing with ill-posed cases are given in Sec. IV. 
Finally, a brief discussion on the robustness of our method is provided in Sec. V.

\section{Hamiltonian reconstruction from eigenstates measurements}

Considering a quantum system with Hilbert space dimension $d$, the system Hamiltonian has the form 
\begin{equation}
\label{eq:ham}
H=\sum_i c_i A_i.
\end{equation}
For a many-body system with $N$ qubits, we have $d=2^N$, and $A_i$s are $k$-local operators. 
Since our theory will apply to any \textit{generic} system with Hamiltonian of the form given in Eq.~\eqref{eq:ham}, we will
treat $A_i$ in general forms for presenting our method. It then naturally applies when $A_i$s are $k$-local
operators.

For a quantum state $\rho$ of the system, we measure the operators $A_i$ and return the expectation values
\begin{equation}
a_i=\tr(\rho A_i).
\end{equation}
If the coefficients $c_{i}$s are known for $H$, it is straightforward to find $a_{i}$s for any system.
This then defines a map 
\begin{equation}
\mathcal{F}: \{c_i\}\rightarrow\{a_i\}.
\end{equation}

We would like to know the situations where the values of $a_{i}$s are enough to determine $\rho$. That is, the situation the inverse problem
\begin{equation}
\mathcal{F}^{-1}: \{a_i\}\rightarrow\{c_i\}
\end{equation}
is well-posed. 
It is known that for any thermal state
\begin{equation}
\rho_{\beta}=\frac{e^{-\beta H}}{\tr e^{-\beta H}},
\end{equation}
for temperature $T$, where $\beta=1/kT$, $\mathcal{F}^{-1}$ is in fact well defined and unique~\cite{cao2020supervised}. That is, the $a_{i}$s uniquely determine $c_{i}$s, hence determine $H$ and $\rho$.

For an eigenstate $\ket{\psi}$, i.e.
\begin{equation}
H\ket{\psi}=\lambda\ket{\psi},
\end{equation}
and $a_i=\tr(\ket{\psi}\bra{\psi} A_i)$,
in general there are many states $\rho$ with $\tr(\rho A_i)=a_i$.
However, restricted to the case that $\rho$ must be an eigenstate of $H$,
there is only one $\rho$ that returns $\tr(\rho A_i)=a_i$
(i.e. $\rho=\ket{\psi}\bra{\psi}$) for most of the cases~\cite{qi2019determining, hou2020determining}. Namely, the map $\mathcal{F}^{-1}$ is generically well-defined.

\begin{figure*}[!htb]
	
	\subfigure[Expectation values on two $3$-qubit random operators $A_1$ and $A_2$, for eigenstates $|\psi_{n} \rangle$ of $H$, where $n=0,1,2,3$ correspond to the ground state, the first, second and third excited state, respectively.]{
		\begin{minipage}{0.45\linewidth}
			\includegraphics[width=0.8\linewidth]{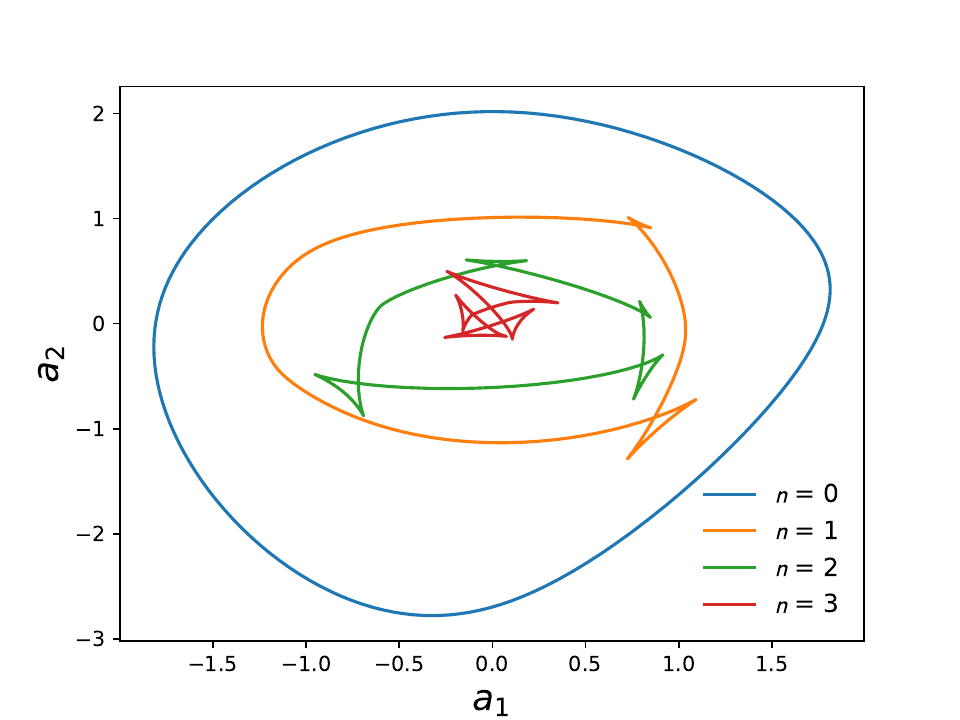}
		\end{minipage}
	}\subfigure[Expectation values on two $3$-qubit random local operators $A_1$ and $A_2$, for eigenstates $|\psi_{n} \rangle$ of $H$, where $n=0,1,2,3$ correspond to the ground state, the first, second and third excited state, respectively.]{
		\begin{minipage}{0.45\linewidth}
			\includegraphics[width=0.8\linewidth]{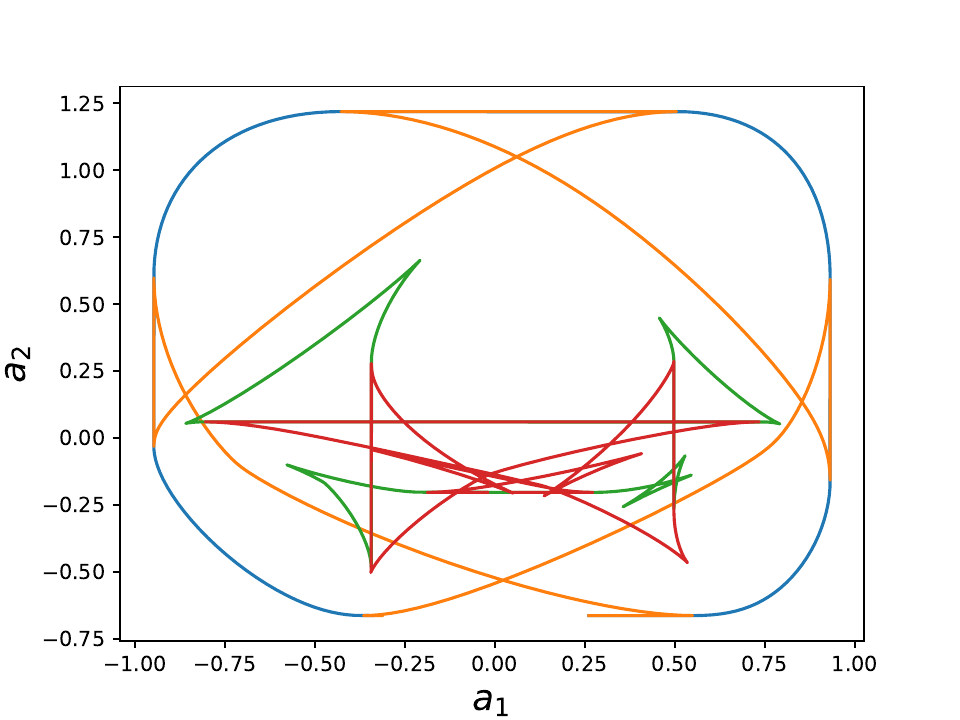}
		\end{minipage}
	}\\
	
	\subfigure[The energy of $H$ in (a) with respect to the eigenstates $|\psi_{n} \rangle$ as a function of $\theta$, for $n=0,\ldots,7$]{
		\begin{minipage}{0.45\linewidth}
			\includegraphics[width=0.8\linewidth]{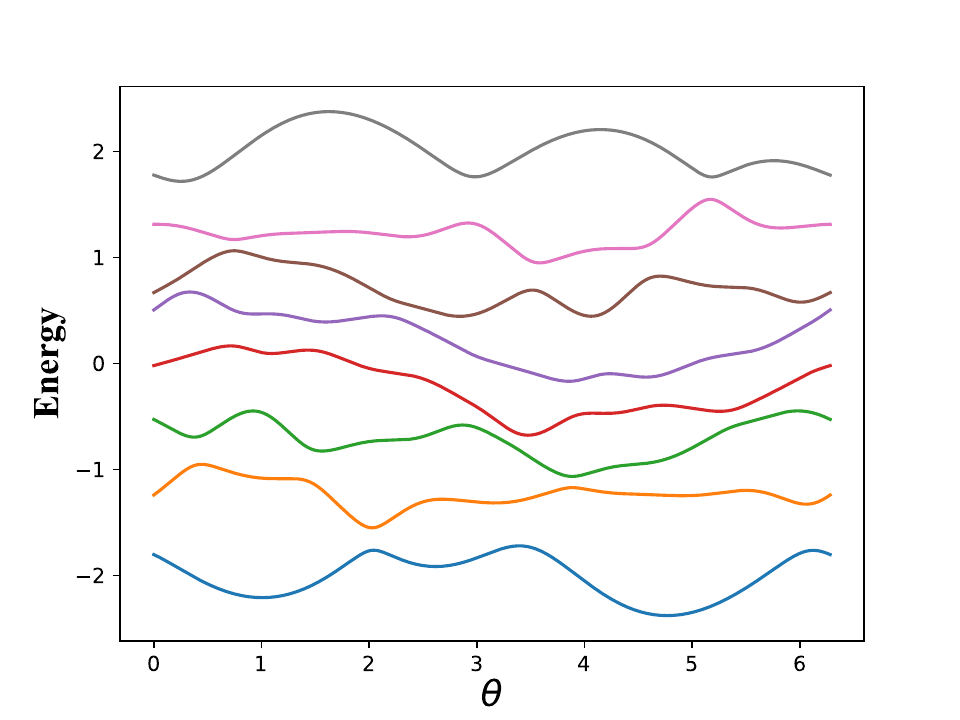}
		\end{minipage}
	}\subfigure[The energy of $H$ in (b) with respect to the eigenstates $|\psi_{n} \rangle$ as a function of $\theta$, for $n=0,\ldots,7$]{
		\begin{minipage}{0.45\linewidth}
			\includegraphics[width=0.8\linewidth]{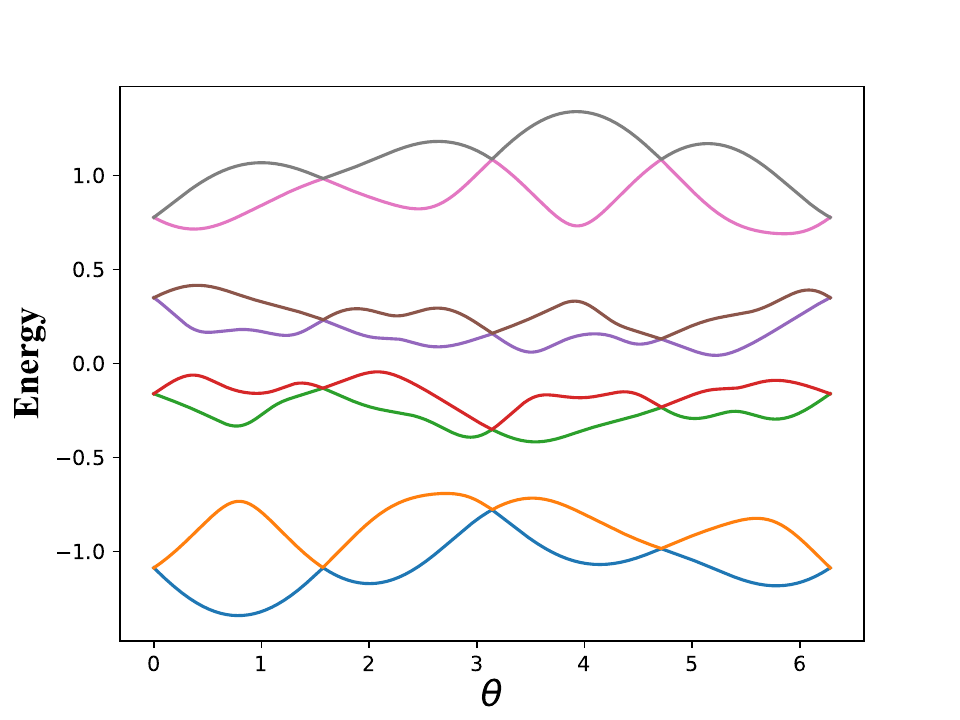}
		\end{minipage}
	}
	\caption{In (a)(c), $A_1$, $A_2$ are nonlocal operators, $a_{i} = \tr(\rho A_i)$, $H = \cos\theta A_{1} +  \sin \theta A_{2}$. In (b)(d), $A_1$, $A_2$ are 2-local operators. $n$ is the level index of $|\psi \rangle$. The blue, orange, green, red, purple, brown, pink and gray curves correspond to the ground state, the 1st, 2nd, 3rd, 4th, 5th, 6th, 7th excited state trajectories (levels) respectively.}
	\label{fig:1}
\end{figure*}

To get an intuition regarding the properties of $\mathcal{F}^{-1}$,
we consider a simple example with $N=3$ qubits.
We generate two random operators $A_{1}$ and $A_{2}$ and set the system Hamiltonian to be 
\begin{equation}
H = \cos\theta A_{1} +  \sin \theta A_{2},
\end{equation}
then choose different eigenstates $|\psi_{n} \rangle$ of $H$ and plot their expectations on $A_1$ and $A_2$ in FIG.~\ref{fig:1}(a). The trajectory of $|\psi_{n} \rangle$ is exactly the same as the trajectory of $|\psi_{2^{3}-1-n} \rangle$, therefore we only plot $n = 0, 1, 2, 3$. Here $n=0,1,2,3$ correspond to the ground, the first, the second and the third excited state, respectively. In FIG.~\ref{fig:1}(c), we plot the energy of $H$ with respect to the eigenstates $|\psi_{n} \rangle$ as a function of $\theta$, for $n=0,\ldots,7$.

For the ground state $|\psi_{0} \rangle$, the trajectory (blue line) is a smooth curve without crossing. 
With the increase of the level index $n$, 
the eigenstate trajectories become more changeable.  There are $2$ crossings for the $1$st excited state, $4$ crossings for the $2$nd and $3$rd excited states. These crossing points correspond to the case of $a_{i}$s where the map $\mathcal{F}^{-1}$ corresponds to multiple $\{c_{i}\}$s.
This indicates that the recovery is well-posed when $|\psi_{n} \rangle$ is low-lying (i.e. $n$ is relatively small), ill-posed when $| \psi_{n} \rangle$ is middle-lying.

We remark that in FIG. ~\ref{fig:1}(a), the area inside the blue line (including the blue line itself) corresponds to the so-called joint numerical range of $A_1$ and $A_2$~\cite{horn2012matrix}. The orange, green, red lines, corresponding to the $1$st, $2$nd, and $3$rd excited states, have a close connection with the higher rank joint numerical ranges of $A_1$ and $A_2$, see e.g.~\cite{chien2011boundary}. 

\begin{figure*}[!htb]
	\subfigure[A $3$-qubit quantum chain structure: each circle represents a qubit, each line represents a specific interaction.]{
		\begin{minipage}{0.3\linewidth}
			\includegraphics[width=0.8\linewidth]{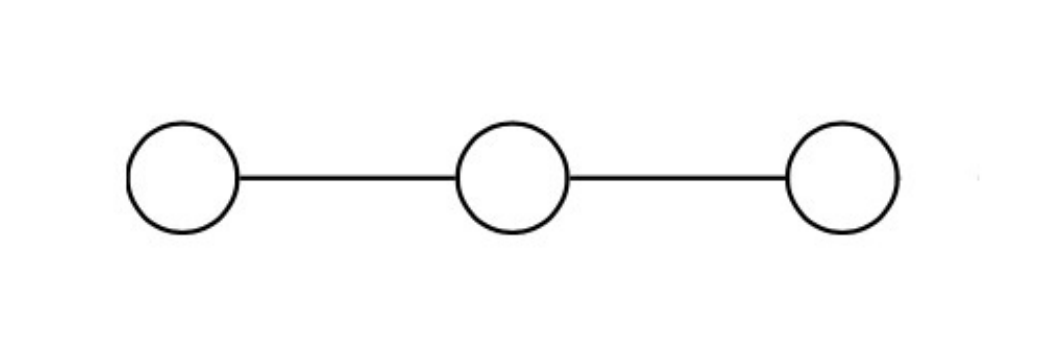}
		\end{minipage}
	}\subfigure[$5$-qubit ring]{
		\begin{minipage}{0.3\linewidth}
			\includegraphics[width=0.9\linewidth]{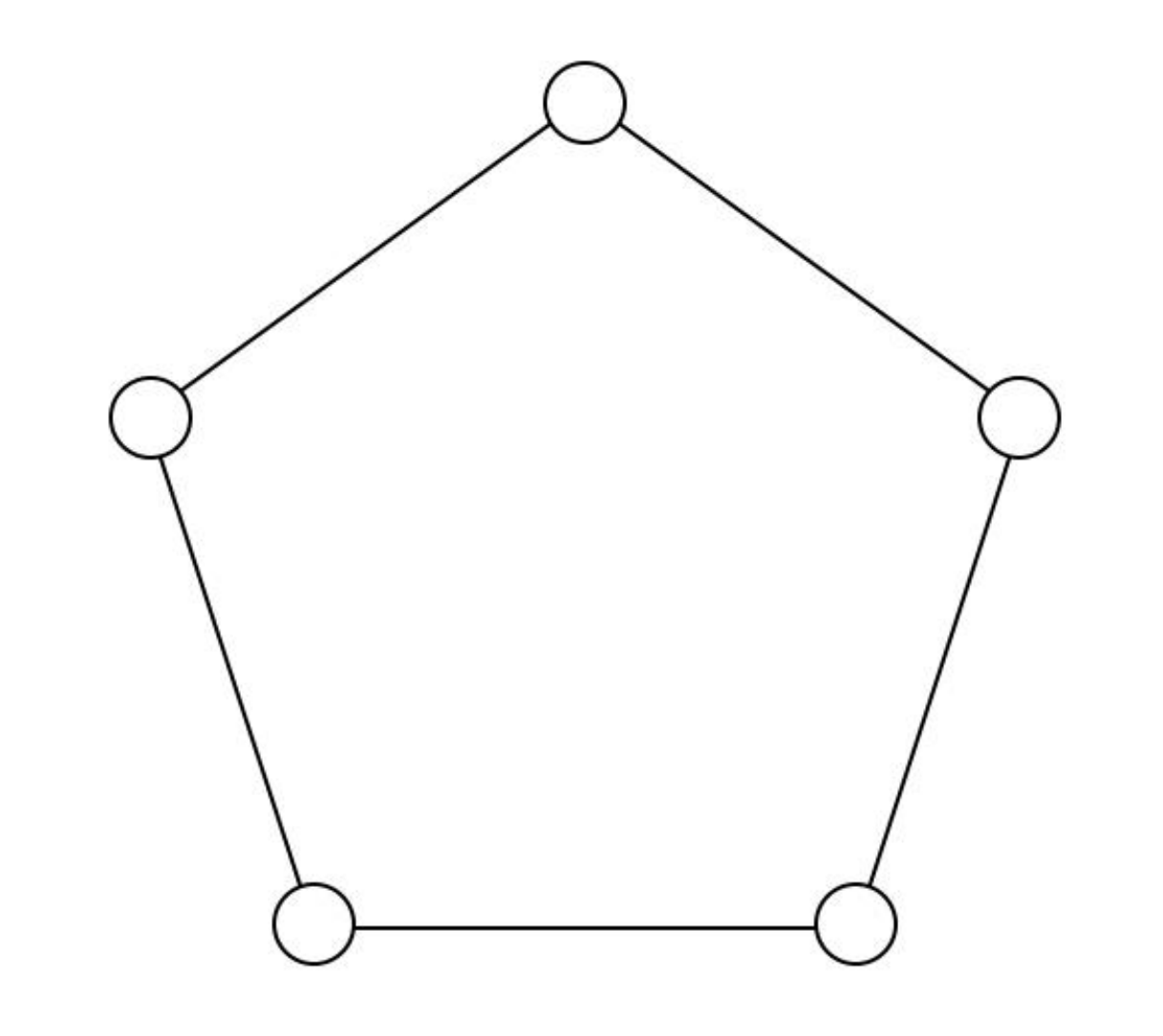}
		\end{minipage}
	}\subfigure[Fully-connected $5$-qubit system]{
		\begin{minipage}{0.3\linewidth}
			\includegraphics[width=0.9\linewidth]{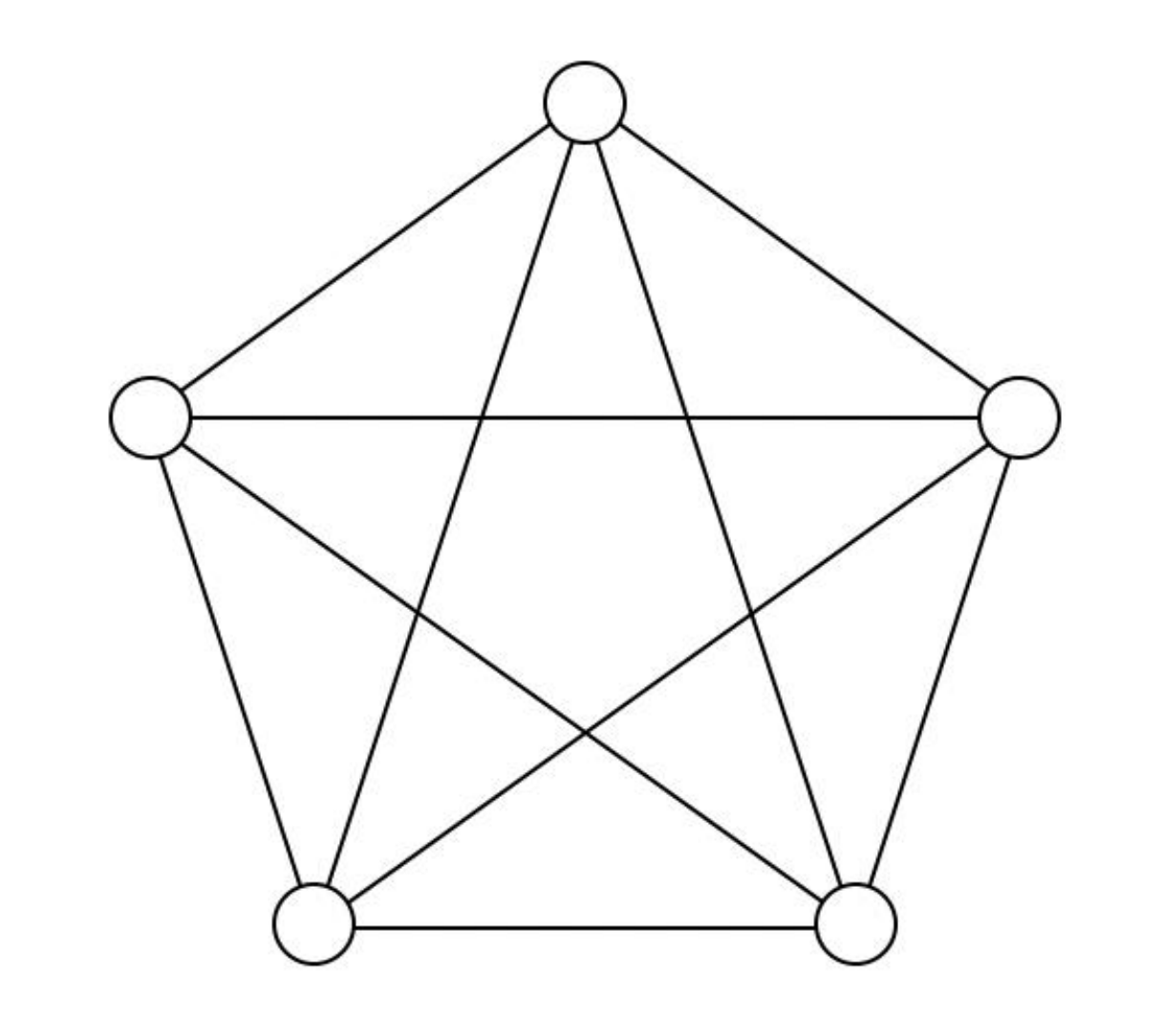}
		\end{minipage}
	}\\
\caption{Some $N$-qubit systems with local structures: a). a line with $3$ qubits; b) a $5$-qubit ring; c) a fully-connected graph of $5$ qubits.}
\label{fig:2}

\end{figure*}

Now we look at a quantum chain with $3$ qubits, as illustrated in FIG.~\ref{fig:2}(a). We choose $2$-local operators $A_1$ and $A_2$ randomly. That is, $A_1$ and $A_2$ act nontrivially only on two neighboring qubits   (the $i$-th qubit and the $(i+1)$-th qubit, for $i=1,2$). Again we choose $H = \cos\theta A_{1} +  \sin \theta A_{2}$. 

We calculate different eigenstates $|\psi_{n} \rangle$ of such $H$ and plot their expectations on $A_1$ and $A_2$ in FIG.~\ref{fig:1}(b).  Here $n=0,1,2,3$ correspond to the ground, the first, the second and the third excited state, respectively. In FIG.~\ref{fig:1}(d), we plot the energy of $H$ with respect to the eigenstates $|\psi_{n} \rangle$ as a function of $\theta$, for $n=0,\ldots,7$. 

In FIG.~\ref{fig:1}(b), there are several ``bridges'' between two paired trajectories, where two eigenstates exchange the expectation values. In FIG.~\ref{fig:1} (d), there are many level crossings between paired energy levels ($\theta = 0, \frac{1}{2} \pi, \pi, \frac{3}{2} \pi, 2 \pi $), which correspond to the ``bridges''. These crossings do not exist for nonlocal energy levels (FIG.~\ref{fig:1}(c)).

By comparison, we observe that recovering local Hamiltonians is more difficult because the solution changes rapidly. This is clear since we cannot use the ``nonlocal" information to distinguish local operators.
In addition, when one qubit is isolated from neighboring qubits, the local Hamiltonian on it is the identity, hence all eigenstates of the system are degenerate.

Although FIG.~\ref{fig:1}(a)(b)(c)(d) are plotted by a single instance of random $A_1$ and $A_2$, we remark that it is not an exotic case. Similar expectation trajectories for some other random $generic$ or random 2-local $A_1$ and $A_2$ are shown in the appendix.

\section{Supervised learning for Hamiltonian reconstruction}

In this section, we discuss our method to reconstruct the system Hamiltonian with supervised learning. We then apply our method to  general $A_i$s and local $A_i$s to reconstruct system Hamiltonians. We further apply our results to catalyze the BFGS method.

\subsection{Method}

Supervised learning is the task of training a parameterized model to match the training set of input-output pairs and make predictions for unseen points~\cite{mohri2018foundations}. The model we use is the artificial neural network. 

A neural network usually contains multiple layers, each layer contains multiple neurons with distinct connections. The leftmost layer is the input layer, and the rightmost one is the output layer. We can have zero or more hidden layers between the input and output layers. The number and size of hidden layers highly influence a neural network's performance, training speed, and convergence. The map between two neighboring layers is a linear transformation followed by a non-linear activation function. Common activation functions include Sigmoid, Tanh, ReLU, Leaky ReLU, Softmax, ELU. A neural network can therefore be regarded as a composition of multiple non-linear functions. We optimize the weights (parameters) in the networks with the backpropagation algorithm to minimize a chosen loss function. 

First, we generate the required local or general operators $\{A_{i}\}$. Since each $2^N$ by $2^N$ Hermitian matrix can be decomposed in the Pauli product basis $\{I, X, Y ,Z\}^{\otimes  N}$ with coefficients $\{x_{P_{1}P_{2}\cdots P_{N}}, P_{i} \in \{I, X, Y, Z\} \}$, we can directly generate a random general $N$-qubit operator $A_{gen}$ by sampling each $x_{P_{1}P_{2}\cdots P_{N}}$ from the interval $[-1, 1]$. 
\begin{equation}
A_{gen} = \sum_{P_{1}, P_{2}, \cdots , P_{n}} x_{P_{1}P_{2}\cdots P_{N}}P_{1} \otimes P_{2} \otimes\cdots  \otimes P_{N}
\end{equation}

We then construct operators with local structures. Consider an $N$-qubit local Hamiltonian with only correlations between the $\alpha$th qubit and $\beta$th qubit, the coefficients $\{x_{I\cdots P_{\alpha}\cdots P_{\beta}\cdots I}\}$ are chosen from the interval $[-1, 1]$ for the corresponding Pauli products $\{ I \otimes  \cdots  \otimes P_{\alpha} \otimes \cdots \otimes P_{\beta} \otimes \cdots \otimes I \}$, the remaining coefficients are set to be 0.  

\begin{equation}
A_{\alpha, \beta} = \sum_{P_{\alpha}, P_{\beta}} x_{I\cdots P_{\alpha}\cdots P_{\beta}\cdots I}I \otimes  \cdots \otimes P_{\alpha} \otimes \cdots  \otimes P_{\beta}  \otimes \cdots  \otimes I 
\end{equation}

Given operators $\{A_i\}$, we uniformly sample 1000 sets of $\{c_i\}$ with each $c_i$ uniformly chosen from $[-1, 1]$. Then for each set, we calculate the corresponding Hamiltonian
\begin{equation}
H = \sum_i c_{i} A_{i}
\end{equation}
and the lower half eigenstates $\{ |\psi _{n} \rangle \}_{n = 0, 1, \cdots , 2^{N-1}-1}$. 

For each $|\psi _{n} \rangle $, we calculate its expectation values $\{a_{i}\}$ on $\{A_i\}$, i.e. $\{a_{i}= \bra{\psi _{n}}A_i\ket{\psi _{n}} \}$.  Now we have $1000 \times 2^{N-1}$ data pairs in the training set, the Hamiltonian reconstruction can be regarded as a regression from $\{a_{i}   \}$ to $\{c_i\}$. In principle, we can generate as many training data as we need, but a small data set is already good enough for most levels. The test set and validation set are generated in the same way. 

We use a shallow neural network with two hidden layers to do the regression. The first hidden layer has 64 neurons, the second one has 32 neurons. The activation function is the Leaky ReLU, which is $max(0.1x, x)$. The learning rate is $2 \times 10^{-4}$. The optimizer is Adam, which can escape saddle points efficiently.

A neural network with two hidden layers and appropriate activation functions can approximate any smooth mapping to any accuracy \cite{heaton2013artificial}. In our method, the first hidden layer learns to extract low-level features of the trajectory (e.g., lines), the second layer learns to extract higher-level features (e.g., combinations of these lines, corners). A neural network with more hidden layers can represent more complex functions. Although the performance can be slightly improved if we replace this network with a deeper neural network, a lot of tricks are required to suppress overfitting and the vanishing gradient problem~\cite{goodfellow2016deep}, much more data and longer time are required for training, the final performance heavily depends on the hyperparameters, these will make the method less practical and universal. Therefore, we use the 2-hidden-layer structure, fix the number of neurons in each layer by numerical tests. Our choice of hyperparameters is a tradeoff between converging rate and performance, 2000 epochs usually suffice.

Denote $\mathbf{c} = (c_{1}, c_{2}, \cdots , c_{p})$, the loss function between the real $\mathbf{c}$ and predicted output $\mathbf{c}'$ is the CosineEmbeddingLoss:

\begin{equation}
\operatorname{loss}(\mathbf{c}, \mathbf{c}')= 1 - \cos (\mathbf{c}, \mathbf{c}')
\end{equation}

The fidelity between the recovered Hamiltonian $H_{rec}$ and the real Hamiltonian $H$ is defined by
\begin{equation}
f\left(H_{rec}, H\right)= \frac{1}{2} + \frac{\operatorname{Tr} H_{rec} H}{2\sqrt{\operatorname{Tr} H_{rec}^{2}} \sqrt{\operatorname{Tr} H^{2}}},
\end{equation}
which is a modification of the fidelity formula in~\cite{fortunato2002design}. 

After adequate training, if we set the network predicted Hamiltonian as the initial point for the optimization algorithm in~\cite{hou2020determining}, we do not need to sample as many initial points as the original method\textemdash the neural network predicted one is quite close to the minimum. Details can be found in the next section.

\subsection{Results}
This section shows the Hamiltonian reconstruction results for general and local operators. Each fidelity in FIG.~\ref{fig:3}(a-c) is averaged by 100 samples. The condition number is a measure of how much the output changes for a small input change. When $|\psi _{n} \rangle $ is a low-lying eigenstate, the condition number is relatively small, the neural networks can recover a system Hamiltonian with high fidelity. We notice that the condition number increases with level index $n$, the performance of neural networks decreases accordingly, but it will not increase with system size $N$. 

In our tests, we generated $2^{N-1} \times 1000$ data pairs for an $N$-qubit system,  1000 pairs for each level.  One can slightly push the average performance of a neural network by using more training data.
However, if we only care about the lowest $n$ energy levels, $n \times 1000$ data pairs will suffice for training. In this case, we do not need to scale the neural network or generate more data points when the system size is increased to more qubits.

\subsubsection{General operators}

Suppose there are $N$ qubits in a system and the Hamiltonian is also given in terms of summation of $N$ general operators $A_{1}, A_{2},\cdots, A_{N}$. 
\begin{equation}
H = \sum_{i=1}^{N} c_{i}A_{i} 
\end{equation}

\begin{figure*}[!btp]
	\centering
	\subfigure[Recovering fidelity versus level index for random Hamiltonian. $f$ is the fidelity between the neural network output Hamiltonian and the real Hamiltonian. Red, green, orange and blue dots correspond to the systems with 5, 6, 7 and 8 qubits respectively. ]{
		\label{fig:subfig:a} 
		\includegraphics[scale=0.5]{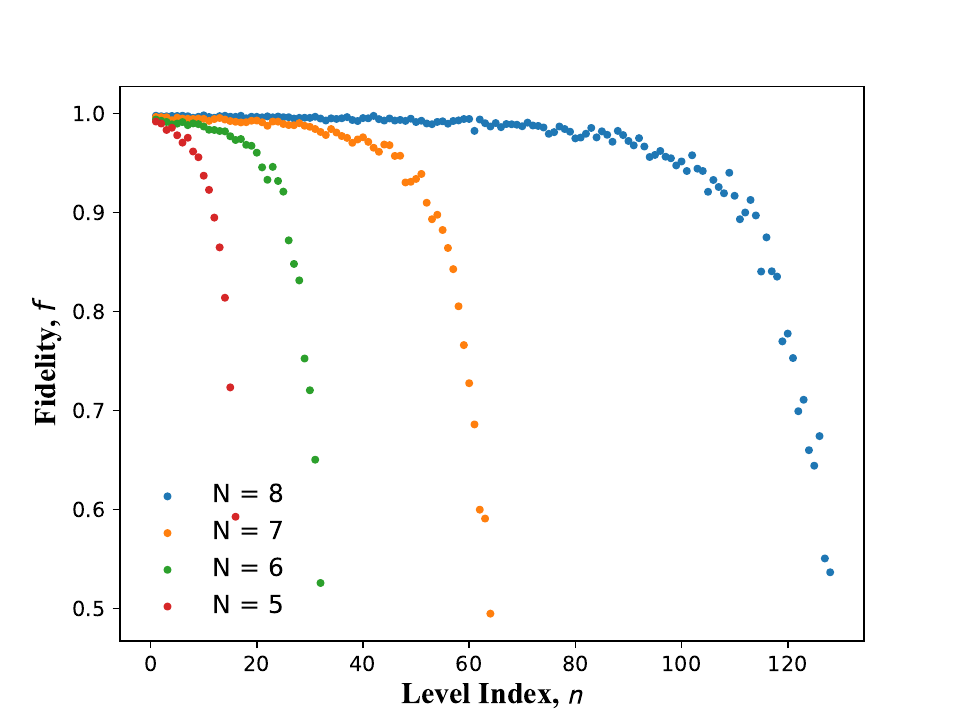}
	}\subfigure[Recovering fidelity versus level index for local ring Hamiltonian. $f$ is the fidelity between the neural network output Hamiltonian and the real Hamiltonian. Red, green, orange and blue dots correspond to the systems with 5, 6, 7 and 8 qubits respectively. ]{
		\label{fig:subfig:b} 
		\includegraphics[scale=0.5]{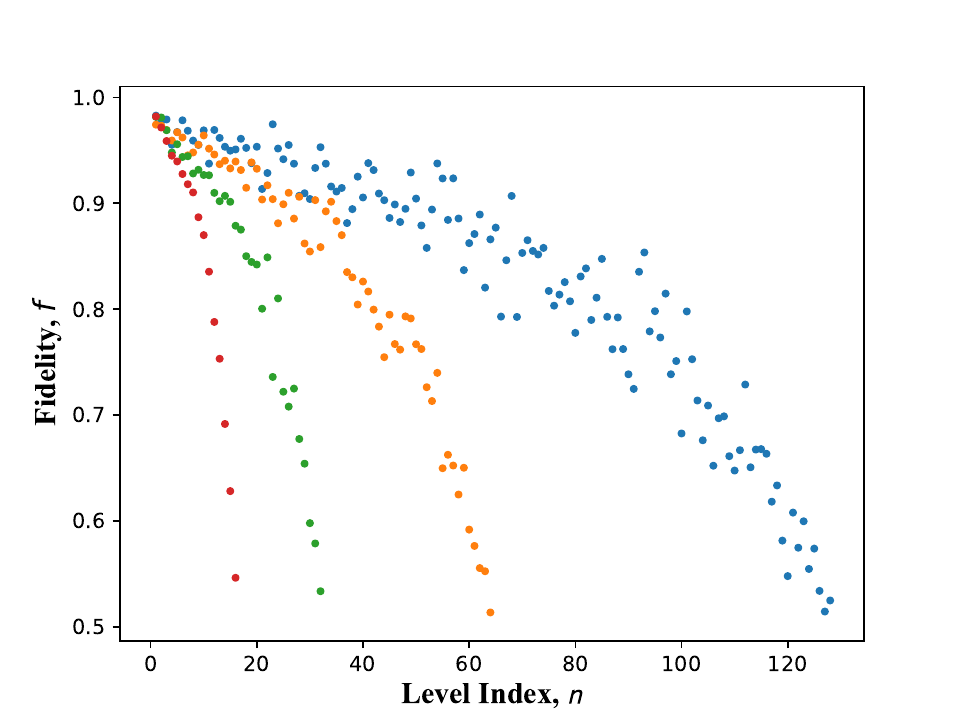}
	}\\
	\subfigure[Recovering fidelity versus level index for local fully-connected Hamiltonian. $f$ is the fidelity between the neural network output Hamiltonian and the real Hamiltonian. Red, green, orange and blue dots correspond to the systems with 5, 6, 7 and 8 qubits respectively. ]{
		\label{fig:subfig:c} 
		\includegraphics[scale=0.5]{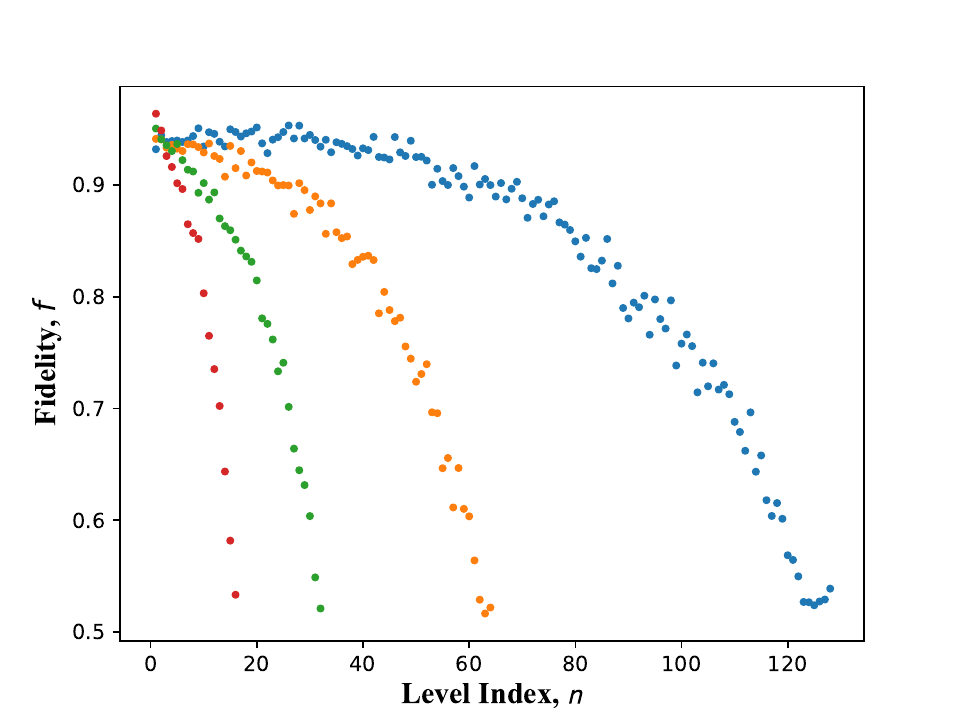}
	}
	\subfigure[Success rate versus level index for different initial points. $f$ is the fidelity between the numerically optimized Hamiltonian and the real Hamiltonian. The solid line is the success rate when we set the Neural Network (NN) predicted Hamiltonian as an initial point for optimization, the dashed line is the success rate when we choose a random initial point.]{
		\label{fig:subfig:d} 
		\includegraphics[scale=0.5]{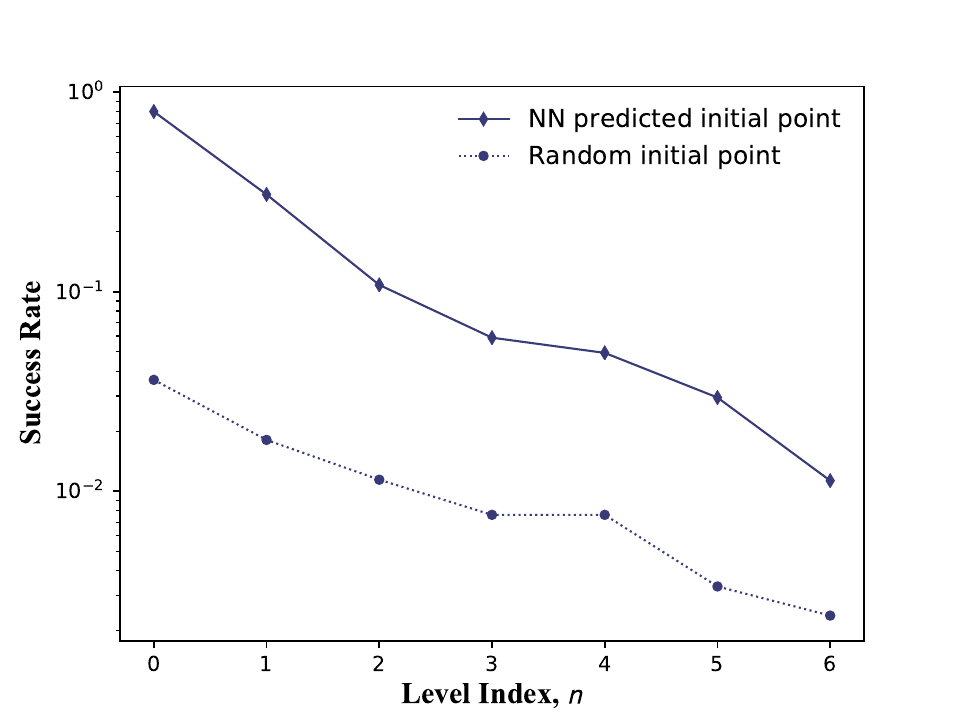}
	}
	\caption{Supervised learning results: (a) General operators; (b) Local operators corresponding to a $5$-qubit ring graph; (c) Local operators corresponding to a $5$-qubit fully-connected graph. (d) Using supervised learning results as initial values for catalyzing the BFGS method.}
	\label{fig:3}
\end{figure*}

We uniformly sample a thousand sets of $\{c_{i} \}$ from the interval $[-1, 1]$, calculate eigenstate expectations $\{a_{i}\}$ for each eigenstate.  Then we train a 3-layer neural network to recover the system Hamiltonian. 

The results are shown in FIG.~\ref{fig:3}(a). The level index $n$ is not input to the network, but can be recorded and traced. Our method performs quite well for most eigenstates, where the fidelities are above $0.99$. The performance will not decrease as the system size increases, indicating scalability. The condition number mainly depends on  ${n}/{2^{N-1}}$

\subsubsection{Local operators}

For local operators, we first consider a $5$-qubit quantum ring, as illustrated in FIG.~\ref{fig:2}(b). There are only interactions between neighboring qubits. The Hamiltonian is
\begin{equation}
H=\sum_{i=1}^{4} c_{i, i+1} A_{i, i+1}+c_{5,1}A_{5,1}
\end{equation}

We apply the same shallow neural network to do the regression. The fidelity decreases rapidly as the level index $n$ increases, as shown in FIG.~\ref{fig:3}(b).  Nevertheless, for the same $n$, the Hamiltonian of a larger system is even easier to reconstruct. The overall performance is worse than that of the general case.

Then we consider $5$-qubit fully-connected systems, as illustrated in FIG.~\ref{fig:2}(c), where the Hamiltonian can be written as 
\begin{equation}
H=\sum_{1 \leq i<j \leq 5} c_{i j} A_{i j}
\end{equation}

We train a $3$-layer neural network to do the regression, and the fidelities are plotted in FIG.~\ref{fig:3}(c). The recovered Hamiltonian is very close to the real one for low-lying eigenstates ($n = 0, 1, 2, 3$). This is good enough since usually we only deal with low-lying eigenstates in experiments. An improved method assisted by transfer learning is discussed in Sec. IV, and it performs much better for middle-lying eigenstates.

\subsection{Initial points for the BFGS method}

Denote the set $\{c_i\}$ by a vector $\mathbf{c}$. In~\cite{hou2020determining}, the BFGS algorithm is used to minimize the objective function 

\begin{equation}
f(\mathbf{x})=\left(\operatorname{tr}\left(A_{i} \rho(\mathbf{x})\right)-a_{i}\right)^{2}+\operatorname{tr}\left(\tilde{H}^{2} \rho(\mathbf{x}),\right)
\end{equation}
where $\mathbf{x}$ is the estimation of $\mathbf{c}$,  $\tilde{H}  = \sum_{i} c_{i} (A_{i}-a_{i} I)$, $\rho(\mathbf{c})=\frac{e^{-\beta \tilde{H}^{2}}}{\operatorname{tr}\left(e^{-\beta \tilde{H}^{2}}\right)}$, $\beta$ is a large constant. This algorithm reconstructs system Hamiltonians with high fidelities for general and local Hamiltonians. However, it is time-consuming and its performance heavily depends on the initial values from Monte Carlo sampling.

This algorithm takes a long time on average to find the solution, due to the fact that the BFGS optimization method may easily be trapped in a local optimal solution. The reconstructed Hamiltonian is therefore not the desired one. Only for a few initial points, we can find the global optimal solution with BFGS optimization and reconstruct the system Hamiltonian with very high fidelity. Given an initial point, we define ``success rate'' as the probability to reconstruct the system Hamiltonian with fidelity $f > 1 - 10^{-8}$. If the initial point of $\mathbf{c}$ is randomly sampled, the success rate is significantly low. Consider the $5$-qubit ring example, the success rate is only $0.036$ for the ground state and $0.018$ for the first excited state. Therefore, a mass of randomly sampled initial points is necessary.

\begin{figure*}[!htbp]
	\subfigure[Constrained expectations on 2-local operators $A_1$, $A_2$. The blue, orange, green, red  curves correspond to the ground state, the 1st, 2nd, 3rd excited state trajectories respectively. $H = \cos\theta A_{1} +  \sin \theta A_{2}$, $\cos\theta > 0, \sin \theta > 0$]{
		\begin{minipage}{0.5\linewidth}
			\includegraphics[width=0.8\linewidth]{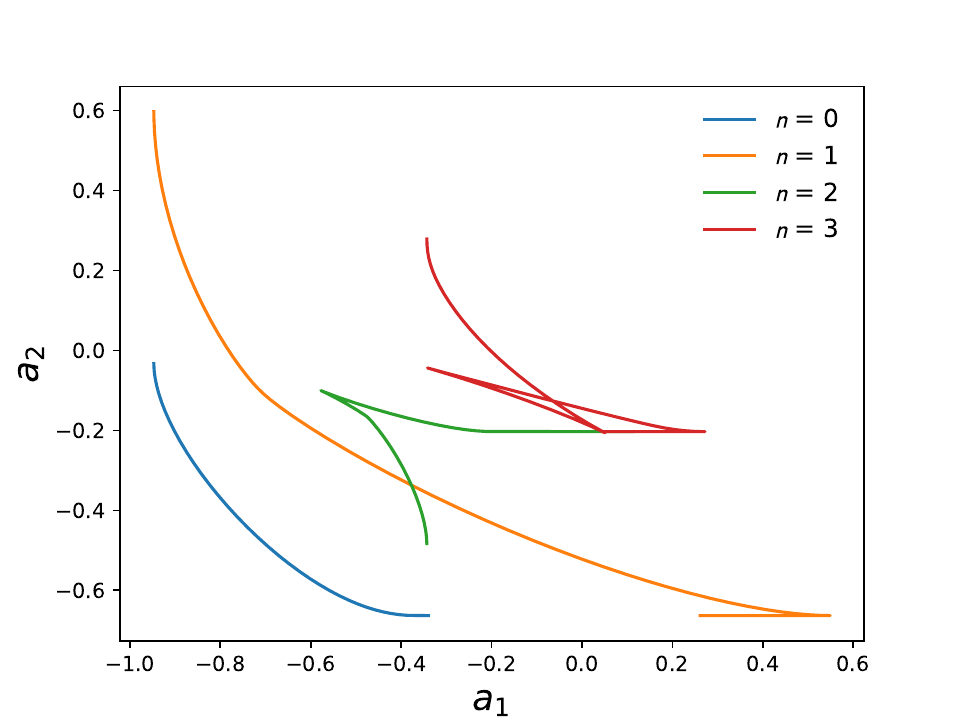}
		\end{minipage}
	}\subfigure[Recovering fidelity versus level index for an N-qubit fully-connected local Hamiltonian assisted by transfer learning. $f$ is the fidelity between the neural network output Hamiltonian and the real Hamiltonian. Red, green, orange and blue dots correspond to the systems with 5, 6, 7 and 8 qubits.]{
		\begin{minipage}{0.5\linewidth}
			\includegraphics[width=0.8\linewidth]{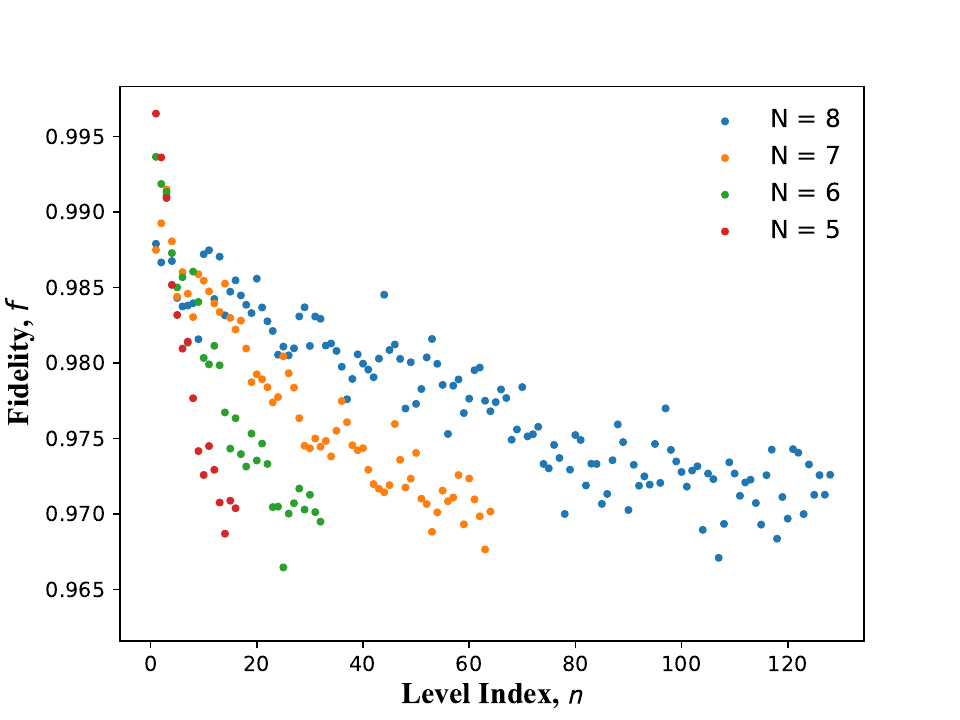}
		\end{minipage}
	}\\
	\caption{A new method for the ill-posed cases.}	
	\label{fig:4}
	
\end{figure*}

However, if we train a neural network and set the predicted Hamiltonian as the initial point for the numerical optimization, the success rate increases dramatically, which is $0.80$ for the ground state and $0.31$ for the first excited state. The results are shown in FIG.~\ref{fig:3}(d), each success rate is calculated by $3000$ samples. The combined algorithm is more efficient than the original one for low-lying energy levels.

\section{Method for the ill-posed cases}

Sometimes, the regression from $\{a_{i}   \}$ to $\{c_i\}$ is not easy for a single neural network because the recovery is ill-posed.  Non-smooth functions, which do not satisfy $f(\boldsymbol{x}+\epsilon \boldsymbol{d}) \approx f(\boldsymbol{x})$ for unit $\boldsymbol{d}$ and small $\epsilon$,  are harder to learn for neural networks \cite{goodfellow2016deep}. However, if there are some constraints on the local Hamiltonian that we can make use of, the sensitivity can be eliminated. 

Take the previous 3-qubit quantum chain as an example, where $H = \cos\theta A_{1} +  \sin \theta A_{2}$. If we restrict $\cos\theta > 0, \sin \theta > 0$, the expectations on $A_1$ and $A_2$ are plotted in FIG.~\ref{fig:4}(a), which is only a small part of FIG.~\ref{fig:1}(b). All break points (energy level crossings) vanish.  The constraints on $\theta$ makes recovery much easier.

Therefore, a multi-class classification network can be used before the regression to preserve smoothness. 

Denote the number of operators as $p$, we divide the generated data to $2^p$ sectors, where each $\{c_i\}$ in the same sector share the same sign. For example, the Hamiltonians of (${c_{1}=\frac{1}{2}, c_{2}=-\frac{1}{2}, c_{3}=\frac{1}{2}}$) and the Hamiltonians of (${c_{1}=\frac{1}{3}, c_{2}=-\frac{1}{3}, c_{3}=\frac{1}{2}}$) belong to the same sector because their signs are the same, which is $(+, -, +)$.

The first neural network will be trained to classify $\{  a_{i}  \}$s to different sectors. It will output a likelihood ranking: the most probable signs of $\{c_i\}$, the second most probable signs of $\{c_i\}$, and so on.

For Hamiltonians in each sector, we divide our generated data into a training set, testing set, and validation set, use a neural network to fit the function from $\{  a_{i}  \}$ to $\{c_i\}$. 

Transfer learning is a machine learning method where the neural network trained in one problem can use as the initial point of a related problem \cite{torrey2010transfer}.  When two Hamiltonians involve the same operators and structure, previously trained weights can be reused to speed up the following training process. For example, the weights in a trained neural network for the $(+, +, +)$ sector can be used as the initial weights for the  $(+, +, -)$ sector. Transfer learning can save us a lot of time for complex neural networks with many hidden layers.

We now summarize our modified algorithm as follows: given operators $\{A_i\}$ and expectations $\{a_{i}\}$, we want to estimate the parameters $\{c_i\}_{\text{est}}$ under these given conditions.


\begin{itemize}
\item Network preparation
\begin{itemize}
	\item [1)] For each sign sector, we sample a thousand sets of $\{c_i\}$ from the corresponding interval, then calculate the Hamiltonian $H = \sum_i c_{i} A_{i}$, eigenstates $\{\ket{\psi _{n}}\}$, and expectations $\{\bra{\psi _{n}} A_{i}\ket{\psi _{n}} \}$ for each set. The total number of data pairs is $2^{p + N-1} \times 1000$.
	\item [2)] Train the multi-class classification network with all sampled data. This network can estimate the sector of $\{c_i\}_{\text{est}}$ (i.e. the signs of $\{c_i\}_{\text{est}}$) for given expectations $\{a_i\}$.
	\item [3)] Train regression networks for each sector with the corresponding $2^{N-1} \times 1000$ data pairs, assisted by transfer learning.
\end{itemize}
\item Estimation
\begin{itemize}
	\item [4)] Input given $\{  a_{i}  \}$ to the classification network and output the sign sector likelihood ranking.
	\item [5)] According to the ranking in 4), input $\{  a_{i}  \}$ to the most probable regression network, verify whether the output Hamiltonian satisfies our requirements. If not, try the next probable regression network.
\end{itemize}
\end{itemize}

The classification neural network we use has three hidden layers, each containing 128 neurons. A feedforward neural network with two or more hidden layers can represent an arbitrary decision boundary to arbitrary accuracy with appropriate activation functions \cite{heaton2013artificial}. The learning rate is $1 \times 10^{-4}$. The activation function is Leaky ReLU. The regression neural network is still the one mentioned in Sec. III A, two hidden layers with 64 and 32 neurons respectively.

For an $N$-qubit system with $2^p$ sectors, we generate $2^{N-1} \times 1000$ data pairs for each sector and train the classification network. After that, we train the regression network for each sector with the same data set. For fully-connected local systems, we implement this method to predict the Hamiltonian, all fidelities are improved remarkably and are above 0.965, as shown in FIG.~\ref{fig:4}(b). The performance of the neural network does not decrease as the number of qubits increases.

If $p$ gets very large, a single classification neural network will not suffice. We can split the operators into several equally-sized parts and train classification neural networks for each part.

\section{Discussion}

In this work, we give a detailed explanation of the Hamiltonian recovery problem and formulate it as an inverse problem.

For the general inverse given by the model $\mathbf{y}=\mathcal{A}\left(x^{*}\right)+\mathbf{e}$,  where $x^{*}$ is the system parameter, $\mathbf{y}$ is the measured data and $\mathbf{e}$ is the observation noise, given data $\mathbf{y}$, we need to recover the model parameter $x^{*}$.  We then propose a supervised learning method via neural networks,
to address the Hamiltonian recovery problem in terms of an inverse problem. Our method can achieve a similar performance with significantly less time than the previous BFGS method, which needs to sample numerous initial points and optimize respectively. Our results demonstrate higher efficiency and can be combined with other optimization algorithms to improve accuracy.

To deal with the error $\mathbf{e}$ in the model, we take the $5$-qubit ring graph as an example. In this case, denote the measurement result $\mathbf{a} = (a_{1}, a_{2}, a_{3}, a_{4}, a_{5})  $. Suppose there is a random error $\delta \mathbf{a}$ during each measurement. First, we generate the data set with the method introduced in Sec. III. After calculating each $\{a_i\}$, we generate a random error $\delta \mathbf{a}$ with a fixed noise ratio $|\delta \mathbf{a}|/|\mathbf{a} |$ and add it to $\{a_i\}$. The measurement result $\{a_i\}$ is therefore inaccurate,  $\mathbf{a}  \rightarrow \mathbf{a} + \delta \mathbf{a}$.

\begin{figure}[!h]
	\includegraphics[scale=0.5]{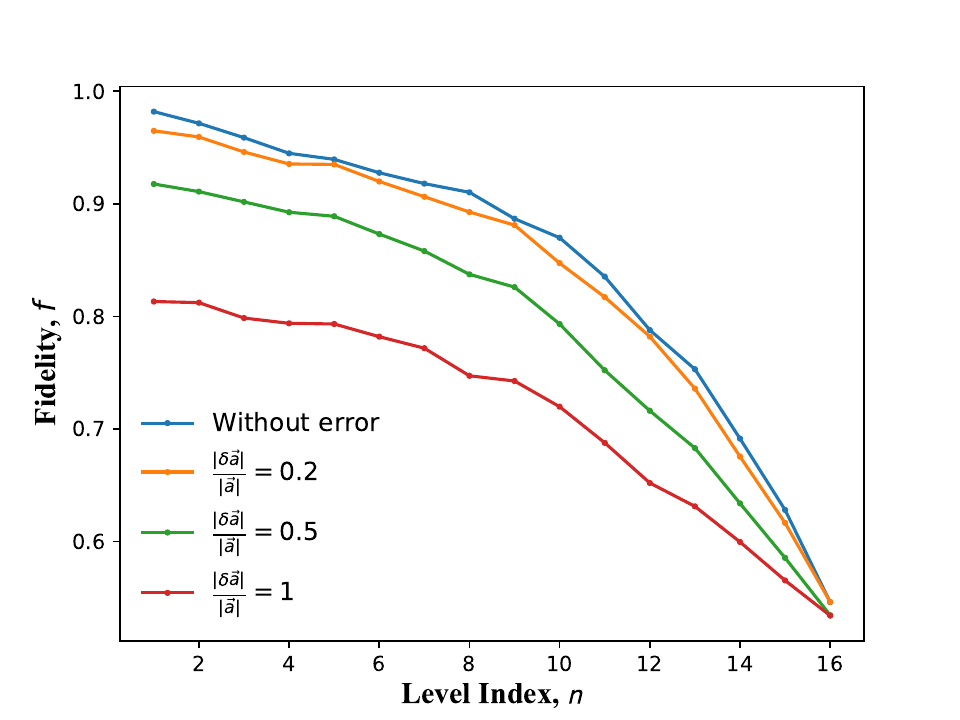}
	\caption{Recovering fidelity versus level index for different errors. $f$ is the fidelity between the neural network output Hamiltonian and the real Hamiltonian. The blue line is the result without errors, the orange line is the result with errors that satisfy $|\delta \mathbf{a}|/|\mathbf{a} | = 0.2$, the green line  corresponds to $|\delta \mathbf{a}|/|\mathbf{a} | = 0.5$, the red line corresponds to $|\delta \mathbf{a}|/|\mathbf{a} | = 1$. }
	\label{fig:5}
\end{figure}

Then we train a 3-layer neural network to predict ${c_i}$s. When the noise is relatively small, i.e. $|\delta \mathbf{a}|/|\mathbf{a} | \leq 0.2$, supervised learning can still recover the system Hamiltonian efficiently, as shown in FIG.~\ref{fig:5}.  The fidelity is close to the noiseless case fidelity for all energy levels.  When $|\delta \mathbf{a}|/|\mathbf{a} | \geq 0.5$, the recovered Hamiltonian is not satisfying, most fidelities are below 0.9.

A small error cannot mix different trajectories due to redundancy in the measurement space. Our method is thus quite robust to errors.

\appendix 
\section{Expectations for 3-qubit operators $A_1$ and $A_2$}

\begin{figure*}[!htb]
	
	\subfigure[]{
		\begin{minipage}{0.33\linewidth}
			\includegraphics[width=1\linewidth]{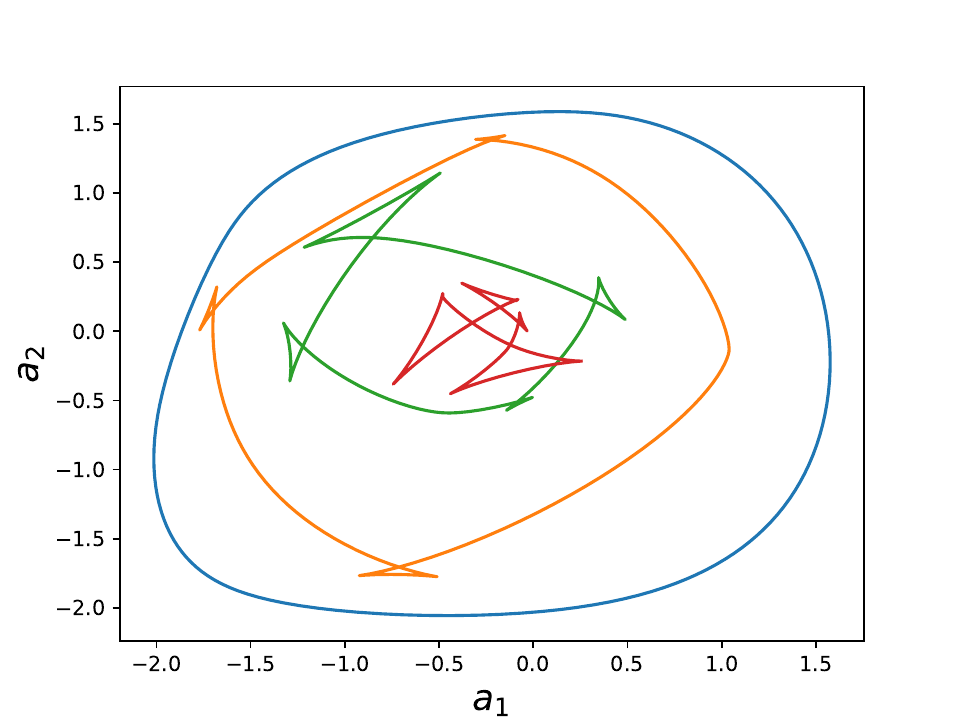}
		\end{minipage}
	}\subfigure[]{
		\begin{minipage}{0.33\linewidth}
			\includegraphics[width=1\linewidth]{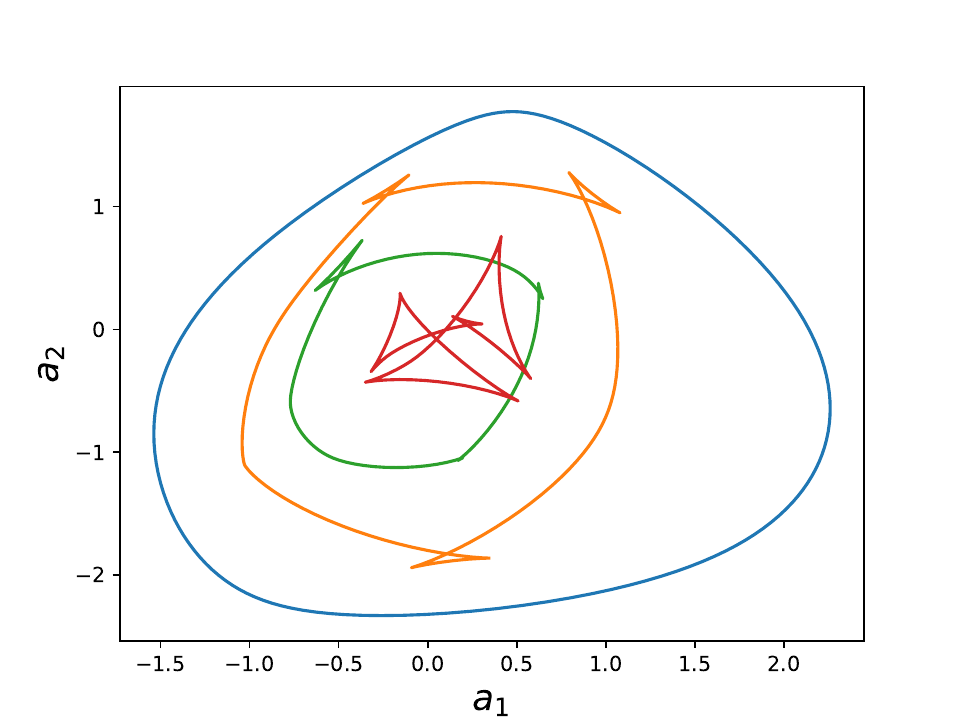}
		\end{minipage}
	}\subfigure[]{
	\begin{minipage}{0.33\linewidth}
		\includegraphics[width=1\linewidth]{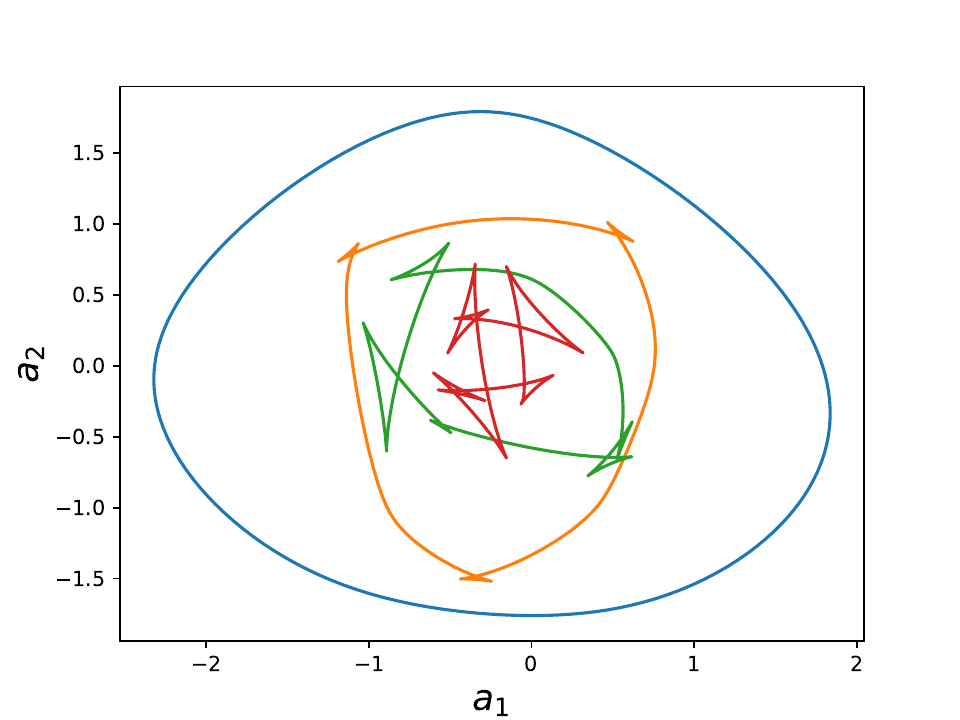}
	\end{minipage}
	}\\
	
	\subfigure[]{
		\begin{minipage}{0.33\linewidth}
			\includegraphics[width=1\linewidth]{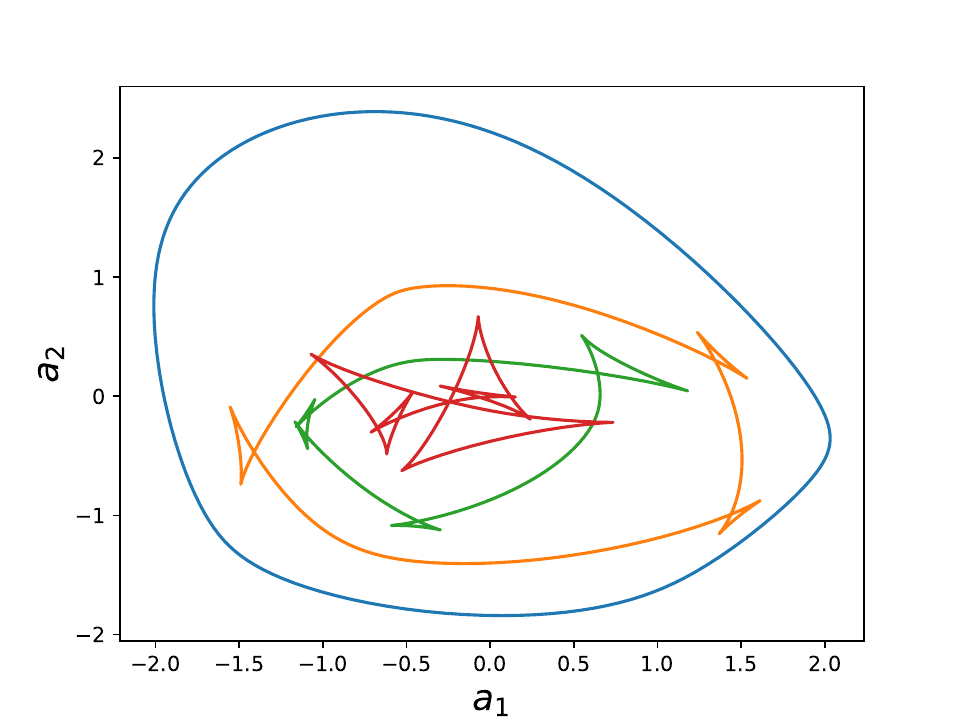}
		\end{minipage}
	}\subfigure[]{
		\begin{minipage}{0.33\linewidth}
			\includegraphics[width=1\linewidth]{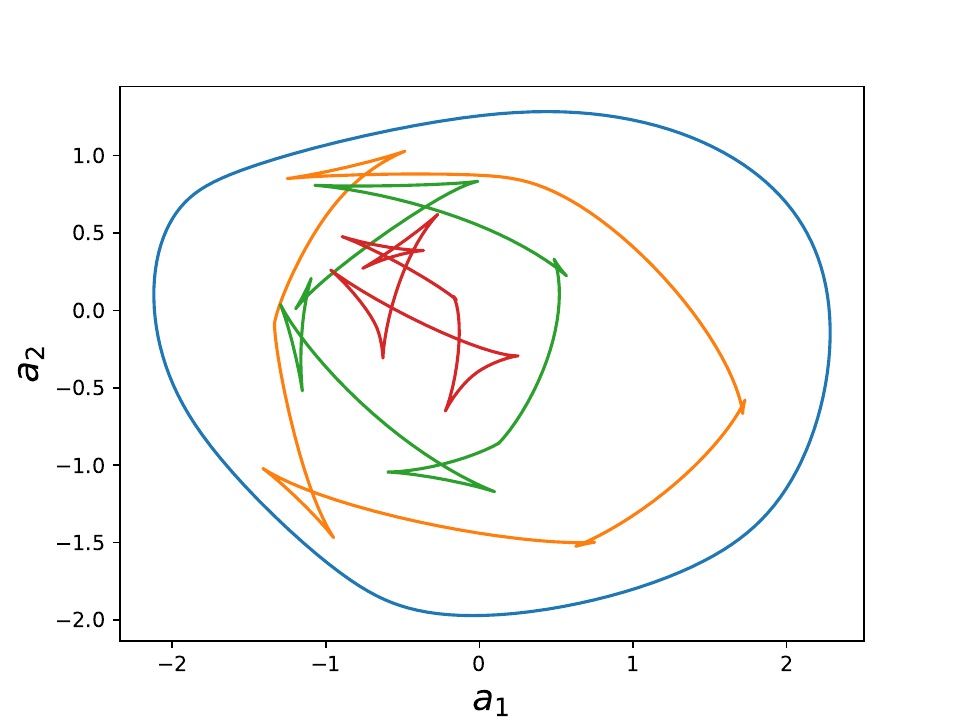}
		\end{minipage}
	}\subfigure[]{
		\begin{minipage}{0.33\linewidth}
			\includegraphics[width=1\linewidth]{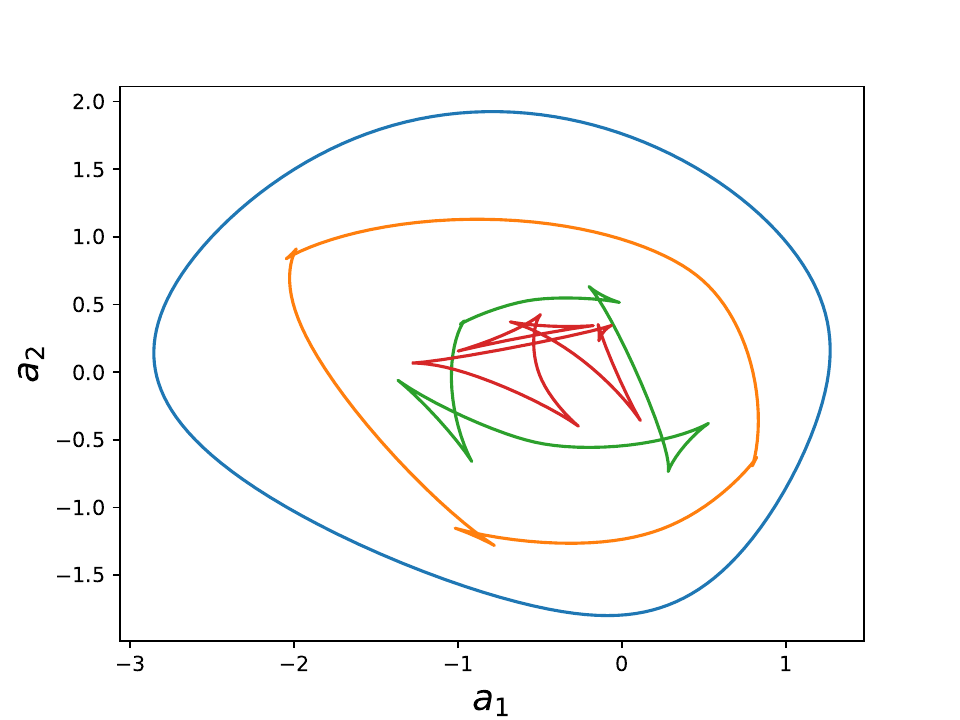}
		\end{minipage}
	}\\

\subfigure[]{
	\begin{minipage}{0.33\linewidth}
		\includegraphics[width=1\linewidth]{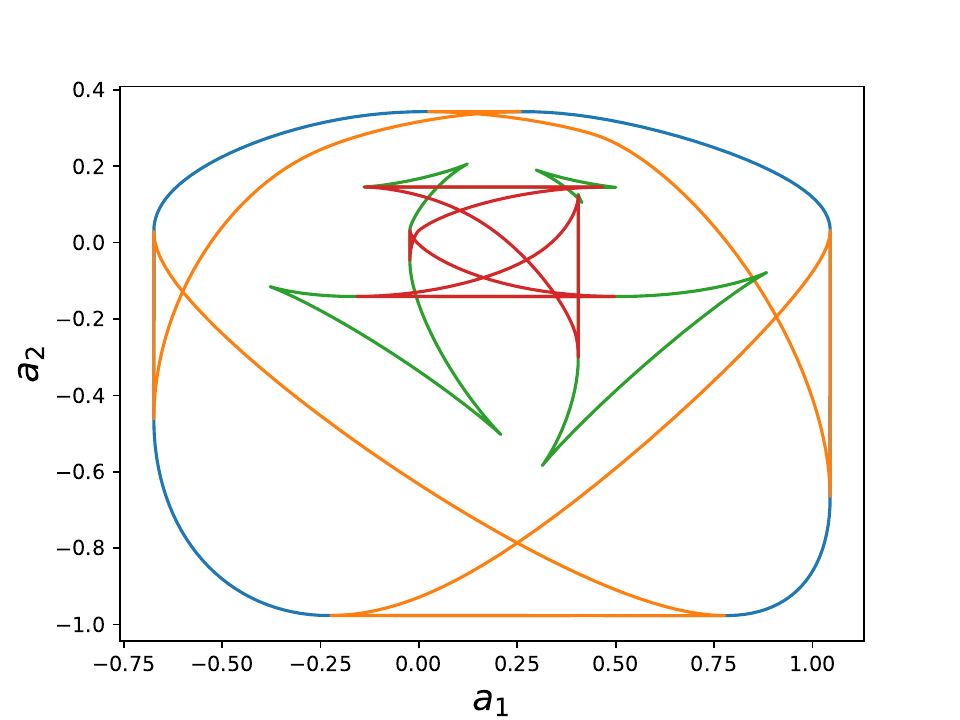}
	\end{minipage}
}\subfigure[]{
	\begin{minipage}{0.33\linewidth}
		\includegraphics[width=1\linewidth]{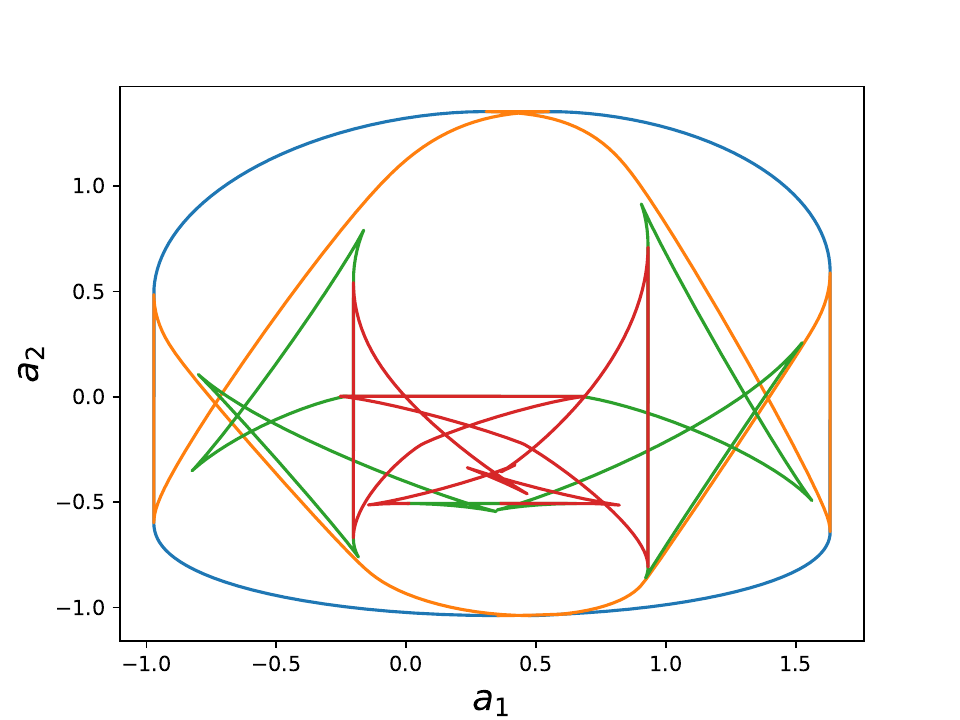}
	\end{minipage}
}\subfigure[]{
	\begin{minipage}{0.33\linewidth}
		\includegraphics[width=1\linewidth]{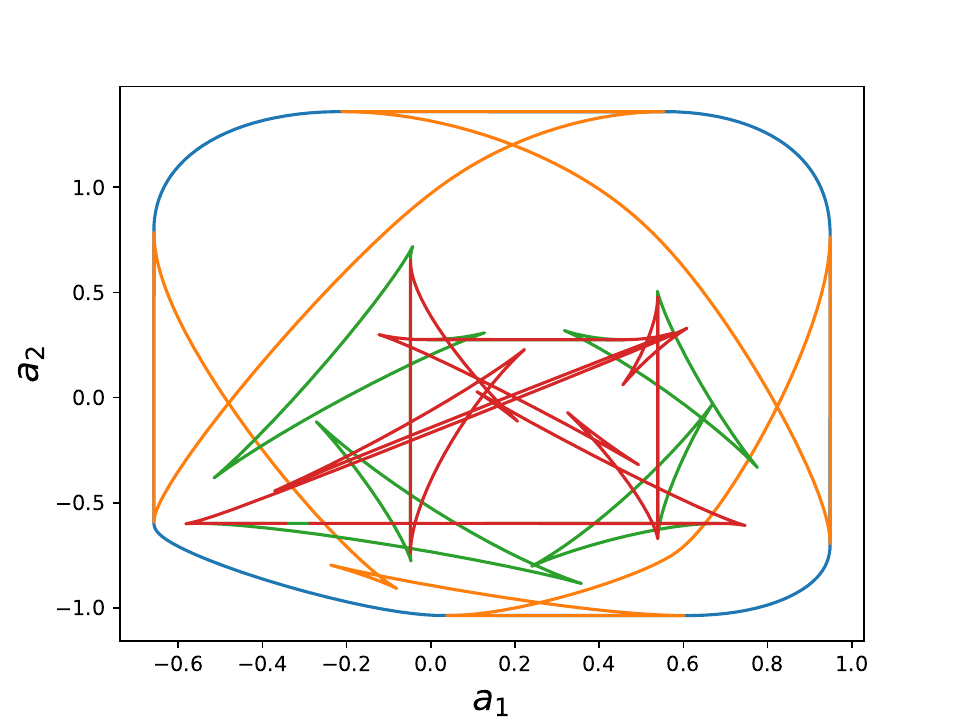}
	\end{minipage}
}\\

\subfigure[]{
	\begin{minipage}{0.33\linewidth}
		\includegraphics[width=1\linewidth]{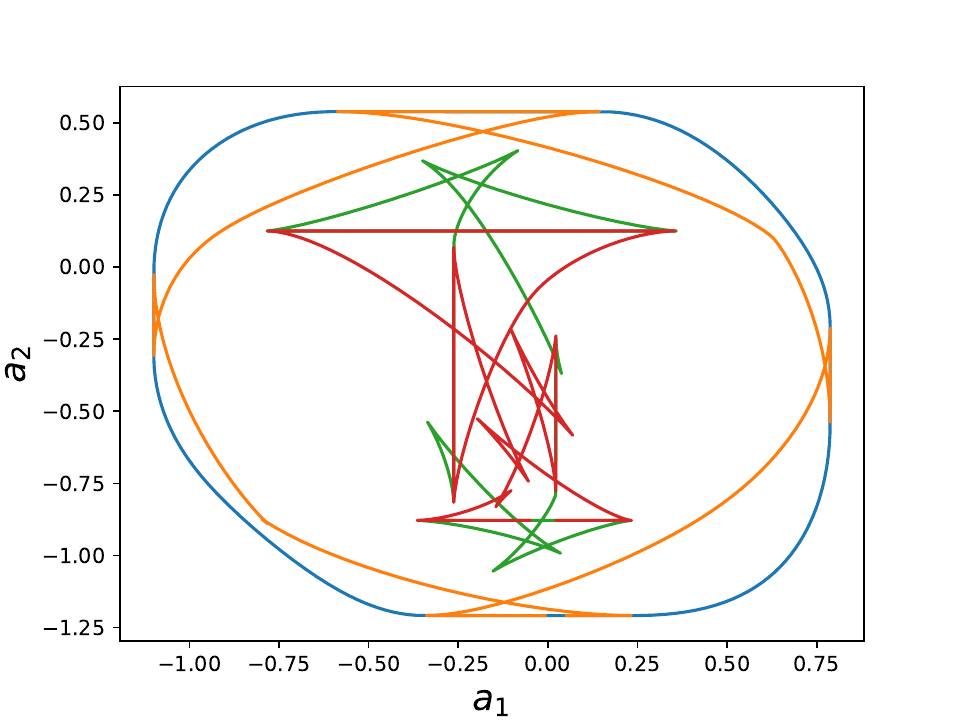}
	\end{minipage}
}\subfigure[]{
	\begin{minipage}{0.33\linewidth}
		\includegraphics[width=1\linewidth]{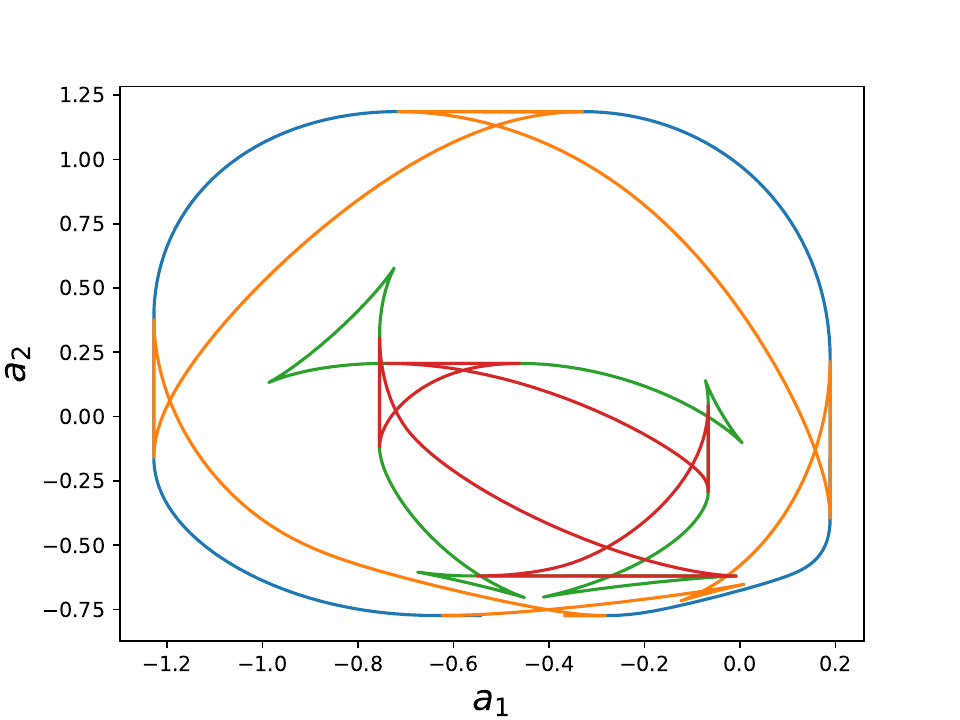}
	\end{minipage}
}\subfigure[]{
	\begin{minipage}{0.33\linewidth}
		\includegraphics[width=1\linewidth]{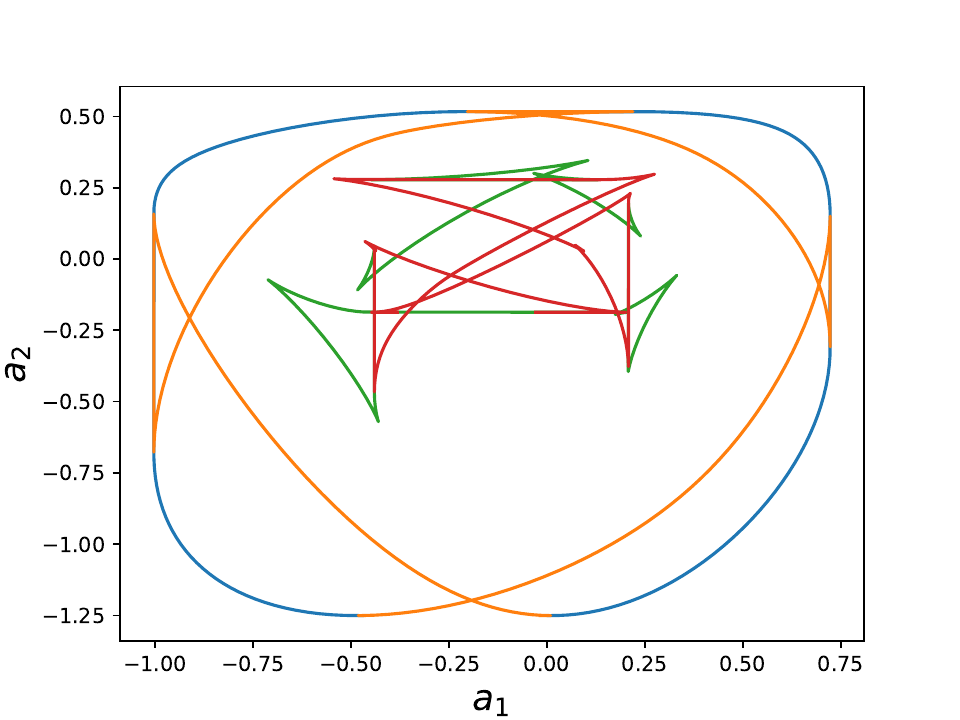}
	\end{minipage}
}
	\caption{Expectation values on two $3$-qubit operators $A_1$ and $A_2$, $a_{i} = \tr(\rho A_i)$, $H = \cos\theta A_{1} +  \sin \theta A_{2}$. In (a)-(f), $A_1$, $A_2$ are nonlocal operators, in (g)-(l), $A_1$, $A_2$ are 2-local operators. The blue, orange, green, red curves correspond to the ground state, the 1st, 2nd, 3rd excited state trajectories (levels) respectively.}
	\label{fig:rand}
\end{figure*}


\begin{thebibliography}{34}
	\expandafter\ifx\csname natexlab\endcsname\relax\def\natexlab#1{#1}\fi
	\expandafter\ifx\csname bibnamefont\endcsname\relax
	\def\bibnamefont#1{#1}\fi
	\expandafter\ifx\csname bibfnamefont\endcsname\relax
	\def\bibfnamefont#1{#1}\fi
	\expandafter\ifx\csname citenamefont\endcsname\relax
	\def\citenamefont#1{#1}\fi
	\expandafter\ifx\csname url\endcsname\relax
	\def\url#1{\texttt{#1}}\fi
	\expandafter\ifx\csname urlprefix\endcsname\relax\def\urlprefix{URL }\fi
	\providecommand{\bibinfo}[2]{#2}
	\providecommand{\eprint}[2][]{\url{#2}}
	
	\bibitem[{\citenamefont{Zeng et~al.}(2015)\citenamefont{Zeng, Chen, Zhou, and
			Wen}}]{zeng2015quantum}
	\bibinfo{author}{\bibfnamefont{B.}~\bibnamefont{Zeng}},
	\bibinfo{author}{\bibfnamefont{X.}~\bibnamefont{Chen}},
	\bibinfo{author}{\bibfnamefont{D.-L.} \bibnamefont{Zhou}}, \bibnamefont{and}
	\bibinfo{author}{\bibfnamefont{X.-G.} \bibnamefont{Wen}},
	\bibinfo{journal}{arXiv preprint arXiv:1508.02595}  (\bibinfo{year}{2015}).
	
	\bibitem[{\citenamefont{Swingle and Kim}(2014)}]{swingle2014reconstructing}
	\bibinfo{author}{\bibfnamefont{B.}~\bibnamefont{Swingle}} \bibnamefont{and}
	\bibinfo{author}{\bibfnamefont{I.~H.} \bibnamefont{Kim}},
	\bibinfo{journal}{Physical review letters} \textbf{\bibinfo{volume}{113}},
	\bibinfo{pages}{260501} (\bibinfo{year}{2014}).
	
	\bibitem[{\citenamefont{Chen et~al.}(2012)\citenamefont{Chen, Ji, Wei, and
			Zeng}}]{chen2012correlations}
	\bibinfo{author}{\bibfnamefont{J.}~\bibnamefont{Chen}},
	\bibinfo{author}{\bibfnamefont{Z.}~\bibnamefont{Ji}},
	\bibinfo{author}{\bibfnamefont{Z.}~\bibnamefont{Wei}}, \bibnamefont{and}
	\bibinfo{author}{\bibfnamefont{B.}~\bibnamefont{Zeng}},
	\bibinfo{journal}{Physical Review A} \textbf{\bibinfo{volume}{85}},
	\bibinfo{pages}{040303} (\bibinfo{year}{2012}).
	
	\bibitem[{\citenamefont{Bairey et~al.}(2019)\citenamefont{Bairey, Arad, and
			Lindner}}]{bairey2019learning}
	\bibinfo{author}{\bibfnamefont{E.}~\bibnamefont{Bairey}},
	\bibinfo{author}{\bibfnamefont{I.}~\bibnamefont{Arad}}, \bibnamefont{and}
	\bibinfo{author}{\bibfnamefont{N.~H.} \bibnamefont{Lindner}},
	\bibinfo{journal}{Physical review letters} \textbf{\bibinfo{volume}{122}},
	\bibinfo{pages}{020504} (\bibinfo{year}{2019}).
	
	\bibitem[{\citenamefont{Qi and Ranard}(2019)}]{qi2019determining}
	\bibinfo{author}{\bibfnamefont{X.-L.} \bibnamefont{Qi}} \bibnamefont{and}
	\bibinfo{author}{\bibfnamefont{D.}~\bibnamefont{Ranard}},
	\bibinfo{journal}{Quantum} \textbf{\bibinfo{volume}{3}}, \bibinfo{pages}{159}
	(\bibinfo{year}{2019}).
	
	\bibitem[{\citenamefont{Linden et~al.}(2002)\citenamefont{Linden, Popescu, and
			Wootters}}]{linden2002almost}
	\bibinfo{author}{\bibfnamefont{N.}~\bibnamefont{Linden}},
	\bibinfo{author}{\bibfnamefont{S.}~\bibnamefont{Popescu}}, \bibnamefont{and}
	\bibinfo{author}{\bibfnamefont{W.}~\bibnamefont{Wootters}},
	\bibinfo{journal}{Physical review letters} \textbf{\bibinfo{volume}{89}},
	\bibinfo{pages}{207901} (\bibinfo{year}{2002}).
	
	\bibitem[{\citenamefont{Baldwin et~al.}(2016)\citenamefont{Baldwin, Deutsch,
			and Kalev}}]{baldwin2016strictly}
	\bibinfo{author}{\bibfnamefont{C.~H.} \bibnamefont{Baldwin}},
	\bibinfo{author}{\bibfnamefont{I.~H.} \bibnamefont{Deutsch}},
	\bibnamefont{and} \bibinfo{author}{\bibfnamefont{A.}~\bibnamefont{Kalev}},
	\bibinfo{journal}{Physical Review A} \textbf{\bibinfo{volume}{93}},
	\bibinfo{pages}{052105} (\bibinfo{year}{2016}).
	
	\bibitem[{\citenamefont{Huang et~al.}(2018)\citenamefont{Huang, Chen, Li, and
			Zeng}}]{huang2018quantum}
	\bibinfo{author}{\bibfnamefont{S.}~\bibnamefont{Huang}},
	\bibinfo{author}{\bibfnamefont{J.}~\bibnamefont{Chen}},
	\bibinfo{author}{\bibfnamefont{Y.}~\bibnamefont{Li}}, \bibnamefont{and}
	\bibinfo{author}{\bibfnamefont{B.}~\bibnamefont{Zeng}},
	\bibinfo{journal}{SCIENCE CHINA Physics, Mechanics \& Astronomy}
	\textbf{\bibinfo{volume}{61}}, \bibinfo{pages}{110311}
	(\bibinfo{year}{2018}).
	
	\bibitem[{\citenamefont{Karuvade et~al.}(2018)\citenamefont{Karuvade, Johnson,
			Ticozzi, and Viola}}]{karuvade2018generic}
	\bibinfo{author}{\bibfnamefont{S.}~\bibnamefont{Karuvade}},
	\bibinfo{author}{\bibfnamefont{P.~D.} \bibnamefont{Johnson}},
	\bibinfo{author}{\bibfnamefont{F.}~\bibnamefont{Ticozzi}}, \bibnamefont{and}
	\bibinfo{author}{\bibfnamefont{L.}~\bibnamefont{Viola}},
	\bibinfo{journal}{Journal of Physics A: Mathematical and Theoretical}
	\textbf{\bibinfo{volume}{51}}, \bibinfo{pages}{145304}
	(\bibinfo{year}{2018}).
	
	\bibitem[{\citenamefont{Zhou}(2008)}]{zhou2008irreducible}
	\bibinfo{author}{\bibfnamefont{D.}~\bibnamefont{Zhou}},
	\bibinfo{journal}{Physical review letters} \textbf{\bibinfo{volume}{101}},
	\bibinfo{pages}{180505} (\bibinfo{year}{2008}).
	
	\bibitem[{\citenamefont{Niekamp et~al.}(2013)\citenamefont{Niekamp, Galla,
			Kleinmann, and G{\"u}hne}}]{niekamp2013computing}
	\bibinfo{author}{\bibfnamefont{S.}~\bibnamefont{Niekamp}},
	\bibinfo{author}{\bibfnamefont{T.}~\bibnamefont{Galla}},
	\bibinfo{author}{\bibfnamefont{M.}~\bibnamefont{Kleinmann}},
	\bibnamefont{and}
	\bibinfo{author}{\bibfnamefont{O.}~\bibnamefont{G{\"u}hne}},
	\bibinfo{journal}{Journal of Physics A: Mathematical and Theoretical}
	\textbf{\bibinfo{volume}{46}}, \bibinfo{pages}{125301}
	(\bibinfo{year}{2013}).
	
	\bibitem[{\citenamefont{Xin et~al.}(2019)\citenamefont{Xin, Lu, Cao, Anikeeva,
			Lu, Li, Long, and Zeng}}]{xin2019local}
	\bibinfo{author}{\bibfnamefont{T.}~\bibnamefont{Xin}},
	\bibinfo{author}{\bibfnamefont{S.}~\bibnamefont{Lu}},
	\bibinfo{author}{\bibfnamefont{N.}~\bibnamefont{Cao}},
	\bibinfo{author}{\bibfnamefont{G.}~\bibnamefont{Anikeeva}},
	\bibinfo{author}{\bibfnamefont{D.}~\bibnamefont{Lu}},
	\bibinfo{author}{\bibfnamefont{J.}~\bibnamefont{Li}},
	\bibinfo{author}{\bibfnamefont{G.}~\bibnamefont{Long}}, \bibnamefont{and}
	\bibinfo{author}{\bibfnamefont{B.}~\bibnamefont{Zeng}}, \bibinfo{journal}{npj
		Quantum Information} \textbf{\bibinfo{volume}{5}}, \bibinfo{pages}{1}
	(\bibinfo{year}{2019}).
	
	\bibitem[{\citenamefont{Cioslowski}(2000)}]{cioslowski2000many}
	\bibinfo{author}{\bibfnamefont{J.}~\bibnamefont{Cioslowski}},
	\emph{\bibinfo{title}{Many-electron densities and reduced density matrices}}
	(\bibinfo{publisher}{Springer Science \& Business Media},
	\bibinfo{year}{2000}).
	
	\bibitem[{\citenamefont{Mazziotti}(1998)}]{mazziotti1998contracted}
	\bibinfo{author}{\bibfnamefont{D.~A.} \bibnamefont{Mazziotti}},
	\bibinfo{journal}{Physical Review A} \textbf{\bibinfo{volume}{57}},
	\bibinfo{pages}{4219} (\bibinfo{year}{1998}).
	
	\bibitem[{\citenamefont{Deutsch}(1991)}]{deutsch1991quantum}
	\bibinfo{author}{\bibfnamefont{J.~M.} \bibnamefont{Deutsch}},
	\bibinfo{journal}{Physical Review A} \textbf{\bibinfo{volume}{43}},
	\bibinfo{pages}{2046} (\bibinfo{year}{1991}).
	
	\bibitem[{\citenamefont{Srednicki}(1994)}]{srednicki1994chaos}
	\bibinfo{author}{\bibfnamefont{M.}~\bibnamefont{Srednicki}},
	\bibinfo{journal}{Physical Review E} \textbf{\bibinfo{volume}{50}},
	\bibinfo{pages}{888} (\bibinfo{year}{1994}).
	
	\bibitem[{\citenamefont{Gogolin and Eisert}(2016)}]{gogolin2016equilibration}
	\bibinfo{author}{\bibfnamefont{C.}~\bibnamefont{Gogolin}} \bibnamefont{and}
	\bibinfo{author}{\bibfnamefont{J.}~\bibnamefont{Eisert}},
	\bibinfo{journal}{Reports on Progress in Physics}
	\textbf{\bibinfo{volume}{79}}, \bibinfo{pages}{056001}
	(\bibinfo{year}{2016}).
	
	\bibitem[{\citenamefont{Garrison and Grover}(2018)}]{garrison2018does}
	\bibinfo{author}{\bibfnamefont{J.~R.} \bibnamefont{Garrison}} \bibnamefont{and}
	\bibinfo{author}{\bibfnamefont{T.}~\bibnamefont{Grover}},
	\bibinfo{journal}{Physical Review X} \textbf{\bibinfo{volume}{8}},
	\bibinfo{pages}{021026} (\bibinfo{year}{2018}).
	
	\bibitem[{\citenamefont{Coleman and Yukalov}(2000)}]{coleman2000reduced}
	\bibinfo{author}{\bibfnamefont{A.~J.} \bibnamefont{Coleman}} \bibnamefont{and}
	\bibinfo{author}{\bibfnamefont{V.~I.} \bibnamefont{Yukalov}},
	\emph{\bibinfo{title}{Reduced density matrices: Coulson’s challenge}},
	vol.~\bibinfo{volume}{72} (\bibinfo{publisher}{Springer Science \& Business
		Media}, \bibinfo{year}{2000}).
	
	\bibitem[{\citenamefont{Hou et~al.}(2020)\citenamefont{Hou, Cao, Lu, Shen,
			Poon, and Zeng}}]{hou2020determining}
	\bibinfo{author}{\bibfnamefont{S.-Y.} \bibnamefont{Hou}},
	\bibinfo{author}{\bibfnamefont{N.}~\bibnamefont{Cao}},
	\bibinfo{author}{\bibfnamefont{S.}~\bibnamefont{Lu}},
	\bibinfo{author}{\bibfnamefont{Y.}~\bibnamefont{Shen}},
	\bibinfo{author}{\bibfnamefont{Y.-T.} \bibnamefont{Poon}}, \bibnamefont{and}
	\bibinfo{author}{\bibfnamefont{B.}~\bibnamefont{Zeng}}, \bibinfo{journal}{New
		Journal of Physics}  (\bibinfo{year}{2020}).
	
	\bibitem[{\citenamefont{Bonnans et~al.}(2006)\citenamefont{Bonnans, Gilbert,
			Lemar{\'e}chal, and Sagastiz{\'a}bal}}]{bonnans2006numerical}
	\bibinfo{author}{\bibfnamefont{J.-F.} \bibnamefont{Bonnans}},
	\bibinfo{author}{\bibfnamefont{J.~C.} \bibnamefont{Gilbert}},
	\bibinfo{author}{\bibfnamefont{C.}~\bibnamefont{Lemar{\'e}chal}},
	\bibnamefont{and} \bibinfo{author}{\bibfnamefont{C.~A.}
		\bibnamefont{Sagastiz{\'a}bal}}, \emph{\bibinfo{title}{Numerical
			optimization: theoretical and practical aspects}}
	(\bibinfo{publisher}{Springer Science \& Business Media},
	\bibinfo{year}{2006}).
	
	\bibitem[{\citenamefont{Adler and {\"O}ktem}(2017)}]{adler2017solving}
	\bibinfo{author}{\bibfnamefont{J.}~\bibnamefont{Adler}} \bibnamefont{and}
	\bibinfo{author}{\bibfnamefont{O.}~\bibnamefont{{\"O}ktem}},
	\bibinfo{journal}{Inverse Problems} \textbf{\bibinfo{volume}{33}},
	\bibinfo{pages}{124007} (\bibinfo{year}{2017}).
	
	\bibitem[{\citenamefont{McCann et~al.}(2017)\citenamefont{McCann, Jin, and
			Unser}}]{mccann2017convolutional}
	\bibinfo{author}{\bibfnamefont{M.~T.} \bibnamefont{McCann}},
	\bibinfo{author}{\bibfnamefont{K.~H.} \bibnamefont{Jin}}, \bibnamefont{and}
	\bibinfo{author}{\bibfnamefont{M.}~\bibnamefont{Unser}},
	\bibinfo{journal}{IEEE Signal Processing Magazine}
	\textbf{\bibinfo{volume}{34}}, \bibinfo{pages}{85} (\bibinfo{year}{2017}).
	
	\bibitem[{\citenamefont{Lucas et~al.}(2018)\citenamefont{Lucas, Iliadis,
			Molina, and Katsaggelos}}]{lucas2018using}
	\bibinfo{author}{\bibfnamefont{A.}~\bibnamefont{Lucas}},
	\bibinfo{author}{\bibfnamefont{M.}~\bibnamefont{Iliadis}},
	\bibinfo{author}{\bibfnamefont{R.}~\bibnamefont{Molina}}, \bibnamefont{and}
	\bibinfo{author}{\bibfnamefont{A.~K.} \bibnamefont{Katsaggelos}},
	\bibinfo{journal}{IEEE Signal Processing Magazine}
	\textbf{\bibinfo{volume}{35}}, \bibinfo{pages}{20} (\bibinfo{year}{2018}).
	
	\bibitem[{\citenamefont{Mousavi and Baraniuk}(2017)}]{mousavi2017learning}
	\bibinfo{author}{\bibfnamefont{A.}~\bibnamefont{Mousavi}} \bibnamefont{and}
	\bibinfo{author}{\bibfnamefont{R.~G.} \bibnamefont{Baraniuk}}, in
	\emph{\bibinfo{booktitle}{2017 IEEE international conference on acoustics,
			speech and signal processing (ICASSP)}} (\bibinfo{organization}{IEEE},
	\bibinfo{year}{2017}), pp. \bibinfo{pages}{2272--2276}.
	
	\bibitem[{\citenamefont{Long et~al.}(2017)\citenamefont{Long, Lu, Ma, and
			Dong}}]{long2017pde}
	\bibinfo{author}{\bibfnamefont{Z.}~\bibnamefont{Long}},
	\bibinfo{author}{\bibfnamefont{Y.}~\bibnamefont{Lu}},
	\bibinfo{author}{\bibfnamefont{X.}~\bibnamefont{Ma}}, \bibnamefont{and}
	\bibinfo{author}{\bibfnamefont{B.}~\bibnamefont{Dong}},
	\bibinfo{journal}{arXiv preprint arXiv:1710.09668}  (\bibinfo{year}{2017}).
	
	\bibitem[{\citenamefont{Cao et~al.}(2020)\citenamefont{Cao, Xie, Zhang, Hou,
			Zhang, and Zeng}}]{cao2020supervised}
	\bibinfo{author}{\bibfnamefont{N.}~\bibnamefont{Cao}},
	\bibinfo{author}{\bibfnamefont{J.}~\bibnamefont{Xie}},
	\bibinfo{author}{\bibfnamefont{A.}~\bibnamefont{Zhang}},
	\bibinfo{author}{\bibfnamefont{S.-Y.} \bibnamefont{Hou}},
	\bibinfo{author}{\bibfnamefont{L.}~\bibnamefont{Zhang}}, \bibnamefont{and}
	\bibinfo{author}{\bibfnamefont{B.}~\bibnamefont{Zeng}},
	\bibinfo{journal}{arXiv preprint arXiv:2005.01540}  (\bibinfo{year}{2020}).
	
	\bibitem[{\citenamefont{Horn and Johnson}(2012)}]{horn2012matrix}
	\bibinfo{author}{\bibfnamefont{R.~A.} \bibnamefont{Horn}} \bibnamefont{and}
	\bibinfo{author}{\bibfnamefont{C.~R.} \bibnamefont{Johnson}},
	\emph{\bibinfo{title}{Matrix analysis}} (\bibinfo{publisher}{Cambridge
		university press}, \bibinfo{year}{2012}).
	
	\bibitem[{\citenamefont{Chien and Nakazato}(2011)}]{chien2011boundary}
	\bibinfo{author}{\bibfnamefont{M.-T.} \bibnamefont{Chien}} \bibnamefont{and}
	\bibinfo{author}{\bibfnamefont{H.}~\bibnamefont{Nakazato}},
	\bibinfo{journal}{Linear algebra and its applications}
	\textbf{\bibinfo{volume}{435}}, \bibinfo{pages}{2971} (\bibinfo{year}{2011}).
	
	\bibitem[{\citenamefont{Mohri et~al.}(2018)\citenamefont{Mohri, Rostamizadeh,
			and Talwalkar}}]{mohri2018foundations}
	\bibinfo{author}{\bibfnamefont{M.}~\bibnamefont{Mohri}},
	\bibinfo{author}{\bibfnamefont{A.}~\bibnamefont{Rostamizadeh}},
	\bibnamefont{and}
	\bibinfo{author}{\bibfnamefont{A.}~\bibnamefont{Talwalkar}},
	\emph{\bibinfo{title}{Foundations of machine learning}}
	(\bibinfo{publisher}{MIT press}, \bibinfo{year}{2018}).
	
	\bibitem[{\citenamefont{Heaton}(2013)}]{heaton2013artificial}
	\bibinfo{author}{\bibfnamefont{J.}~\bibnamefont{Heaton}},
	\emph{\bibinfo{title}{Artificial intelligence for humans}}
	(\bibinfo{publisher}{Heaton Research, Incorporated}, \bibinfo{year}{2013}).
	
	\bibitem[{\citenamefont{Goodfellow et~al.}(2016)\citenamefont{Goodfellow,
			Bengio, and Courville}}]{goodfellow2016deep}
	\bibinfo{author}{\bibfnamefont{I.}~\bibnamefont{Goodfellow}},
	\bibinfo{author}{\bibfnamefont{Y.}~\bibnamefont{Bengio}}, \bibnamefont{and}
	\bibinfo{author}{\bibfnamefont{A.}~\bibnamefont{Courville}},
	\emph{\bibinfo{title}{Deep learning}} (\bibinfo{publisher}{MIT press},
	\bibinfo{year}{2016}).
	
	\bibitem[{\citenamefont{Fortunato et~al.}(2002)\citenamefont{Fortunato, Pravia,
			Boulant, Teklemariam, Havel, and Cory}}]{fortunato2002design}
	\bibinfo{author}{\bibfnamefont{E.~M.} \bibnamefont{Fortunato}},
	\bibinfo{author}{\bibfnamefont{M.~A.} \bibnamefont{Pravia}},
	\bibinfo{author}{\bibfnamefont{N.}~\bibnamefont{Boulant}},
	\bibinfo{author}{\bibfnamefont{G.}~\bibnamefont{Teklemariam}},
	\bibinfo{author}{\bibfnamefont{T.~F.} \bibnamefont{Havel}}, \bibnamefont{and}
	\bibinfo{author}{\bibfnamefont{D.~G.} \bibnamefont{Cory}},
	\bibinfo{journal}{The Journal of chemical physics}
	\textbf{\bibinfo{volume}{116}}, \bibinfo{pages}{7599} (\bibinfo{year}{2002}).
	
	\bibitem[{\citenamefont{Torrey and Shavlik}(2010)}]{torrey2010transfer}
	\bibinfo{author}{\bibfnamefont{L.}~\bibnamefont{Torrey}} \bibnamefont{and}
	\bibinfo{author}{\bibfnamefont{J.}~\bibnamefont{Shavlik}}, in
	\emph{\bibinfo{booktitle}{Handbook of research on machine learning
			applications and trends: algorithms, methods, and techniques}}
	(\bibinfo{publisher}{IGI Global}, \bibinfo{year}{2010}), pp.
	\bibinfo{pages}{242--264}.
	
\end{thebibliography}
\end{document}